\documentclass[prc,superscriptaddress,showpacs,onecolumn,notitlepage,nofootinbib,11pt]{revtex4-1} 
\usepackage{amssymb}
\usepackage{amsmath}
\usepackage{color}
\usepackage{graphicx}
\usepackage{soul}

\def\bibsection{\subsection*{References}}

\begin{document}

\title{Probing the strongly driven spin-boson model in a superconducting quantum circuit}

\author{L. Magazz\`u$^\dag$}
\affiliation{Institute of Physics, University of Augsburg, Universit\"atsstra{\ss}e 1, D-86135 Augsburg, Germany}

\author{P. Forn-D\'iaz$^\dag$}
\affiliation{Institute for Quantum Computing, University of Waterloo, Waterloo N2L 3G1, Canada}
\affiliation{Department of Physics and Astronomy, University of Waterloo, Waterloo N2L 3G1, Canada}
\affiliation{Waterloo Institute for Nanotechnology, University of Waterloo, Waterloo N2L 3G1, Canada}
\affiliation{Barcelona Supercomputing Center (BSC), C/ Jordi Girona 29, 08034 Barcelona, Spain}

\author{R. Belyansky}
\affiliation{Institute for Quantum Computing, University of Waterloo, Waterloo N2L 3G1, Canada}
\affiliation{Department of Electrical and Computer Engineering, University of Waterloo, Waterloo N2L 3G1, Canada}

\author{J.-L. Orgiazzi}
\affiliation{Institute for Quantum Computing, University of Waterloo, Waterloo N2L 3G1, Canada}
\affiliation{Waterloo Institute for Nanotechnology, University of Waterloo, Waterloo N2L 3G1, Canada}
\affiliation{Department of Electrical and Computer Engineering, University of Waterloo, Waterloo N2L 3G1, Canada}

\author{M. A. Yurtalan}
\affiliation{Institute for Quantum Computing, University of Waterloo, Waterloo N2L 3G1, Canada}
\affiliation{Waterloo Institute for Nanotechnology, University of Waterloo, Waterloo N2L 3G1, Canada}
\affiliation{Department of Electrical and Computer Engineering, University of Waterloo, Waterloo N2L 3G1, Canada}

\author{M. R. Otto}
\affiliation{Institute for Quantum Computing, University of Waterloo, Waterloo N2L 3G1, Canada}
\affiliation{Department of Physics and Astronomy, University of Waterloo, Waterloo N2L 3G1, Canada}
\affiliation{Waterloo Institute for Nanotechnology, University of Waterloo, Waterloo N2L 3G1, Canada}

\author{A. Lupascu$^{*,\bot}$}
\affiliation{Institute for Quantum Computing, University of Waterloo, Waterloo N2L 3G1, Canada}
\affiliation{Department of Physics and Astronomy, University of Waterloo, Waterloo N2L 3G1, Canada}
\affiliation{Waterloo Institute for Nanotechnology, University of Waterloo, Waterloo N2L 3G1, Canada}

\author{C. M. Wilson$^{*,\bot}$}
\affiliation{Institute for Quantum Computing, University of Waterloo, Waterloo N2L 3G1, Canada}
\affiliation{Department of Electrical and Computer Engineering, University of Waterloo, Waterloo N2L 3G1, Canada}

\author{M. Grifoni$^*$}
\affiliation{Institute for Theoretical Physics, University of
Regensburg, 93040 Regensburg, Germany}

\let\thefootnote\relax\footnote{
$\dag$ These authors contributed equally to this work.\\
$\bot$ The experimental work was a collaboration between the labs led by these researchers.\\
$^*$ adrian.lupascu@uwaterloo.ca; chris.wilson@uwaterloo.ca; milena.grifoni@ur.de;}

\begin{abstract}
Quantum two-level systems interacting with the surroundings are ubiquitous in nature.
The interaction suppresses quantum coherence and forces the system towards a steady state.  Such dissipative processes are captured by the paradigmatic spin-boson model, describing a two-state particle, the ``spin'',  interacting with an environment formed  by harmonic oscillators. A fundamental  question to date  is to what extent intense coherent driving  impacts a strongly dissipative system.
Here we investigate experimentally and theoretically  a superconducting qubit strongly coupled to an electromagnetic environment and subjected to a coherent drive. 
This setup realizes the driven Ohmic spin-boson model.
We show  that the  drive reinforces environmental suppression of quantum coherence, and that a coherent-to-incoherent transition can be achieved by tuning the drive amplitude. An out-of-equilibrium detailed balance relation 
 is demonstrated.
These results advance fundamental understanding of  open quantum systems and bear potential for the design of entangled light-matter states. 

\end{abstract}

\maketitle

\section*{\large I\MakeLowercase{ntroduction}}

The spin-boson model has been prominent for several decades in the study of open quantum systems  \cite{Leggett1987,Weiss2012}. It describes  a two-state quantum system (spin), interacting with its environment. The latter is modeled as a set of harmonic oscillators (bosons) constituting a  so-called heat bath. 
 The dynamical regimes of the spin-boson model at a given finite temperature are essentially dictated by the coupling to the environment and by the low-frequency  behavior of the bath spectrum.  In the strong coupling regime, this model provides an accurate representation of a variety of physical and chemical situations of broad interest,  including incoherent tunneling of bistable defects  in metals \cite{Golding1992} and amorphous systems \cite{Golding1978}, macroscopic quantum tunneling in superconducting circuits \cite{Han1991}, or  electron and proton  transfer in solvent environments  \cite{Morillo1989}. Moreover, the spin-boson model is relevant in describing exciton transport in biological complexes~\cite{Thorwart2009, Huelga2013}.
The weak coupling regime characterizes situations where preserving quantum coherence is crucial, such as in quantum computing, whereas  strong coupling can give rise to novel entangled states of system and reservoir, for example, to polaron or Kondo clouds \cite{Weiss2012}.\\
\noindent  In the Ohmic spin-boson model, the environment has a linear spectrum at low frequencies which leads to various remarkable phenomena, such as bath-induced localization or a coherent-to-incoherent transition even at zero temperature 
for large enough coupling strengths \cite{Leggett1987}. \\
\indent Recently, a new experimental setup was implemented~\cite{Forn-Diaz2017} which realizes the Ohmic spin-boson model with an environmental coupling tunable from weak to ultrastrong~\cite{Peropadre2013}. This particular implementation is formed from a superconducting flux qubit coupled to a  transmission line, which play the role of the two-state system and environment, respectively. The tunability of the interaction allows one to test the key predictions of the spin-boson model. In \cite{Yoshihara2017}, a qubit ultrastrongly coupled to a single oscillator mode was demonstrated.\\
\noindent In this article, we study the spin-boson setup from Ref.~\cite{Forn-Diaz2017} under strong driving, which adds a new dimension of exploration for a spin-boson system \cite{Grifoni1998}. Previous experiments studying strongly driven systems have  reported remarkable effects, such as the formation of dressed states \cite{Nakamura2001,Wilson2007,Wilson2010},  Landau-Zener interference \cite{Oliver2005,Sillanpaa2006},  amplitude spectroscopy \cite{Berns2008}, and the observation of Floquet states \cite{Deng2015}. However,  these experimental reports were restricted to weak or moderate coupling to the environment. Here, we combine  intense driving and diverse dissipation strengths in a superconducting qubit circuit, with the aim of tracing out the dynamical phase diagram of a driven spin-boson system in coupling regimes ranging form weak to  ultrastrong. \\
\section*{\large R\MakeLowercase{esults}}
\begin{figure}
\centering
\includegraphics[width=10cm]{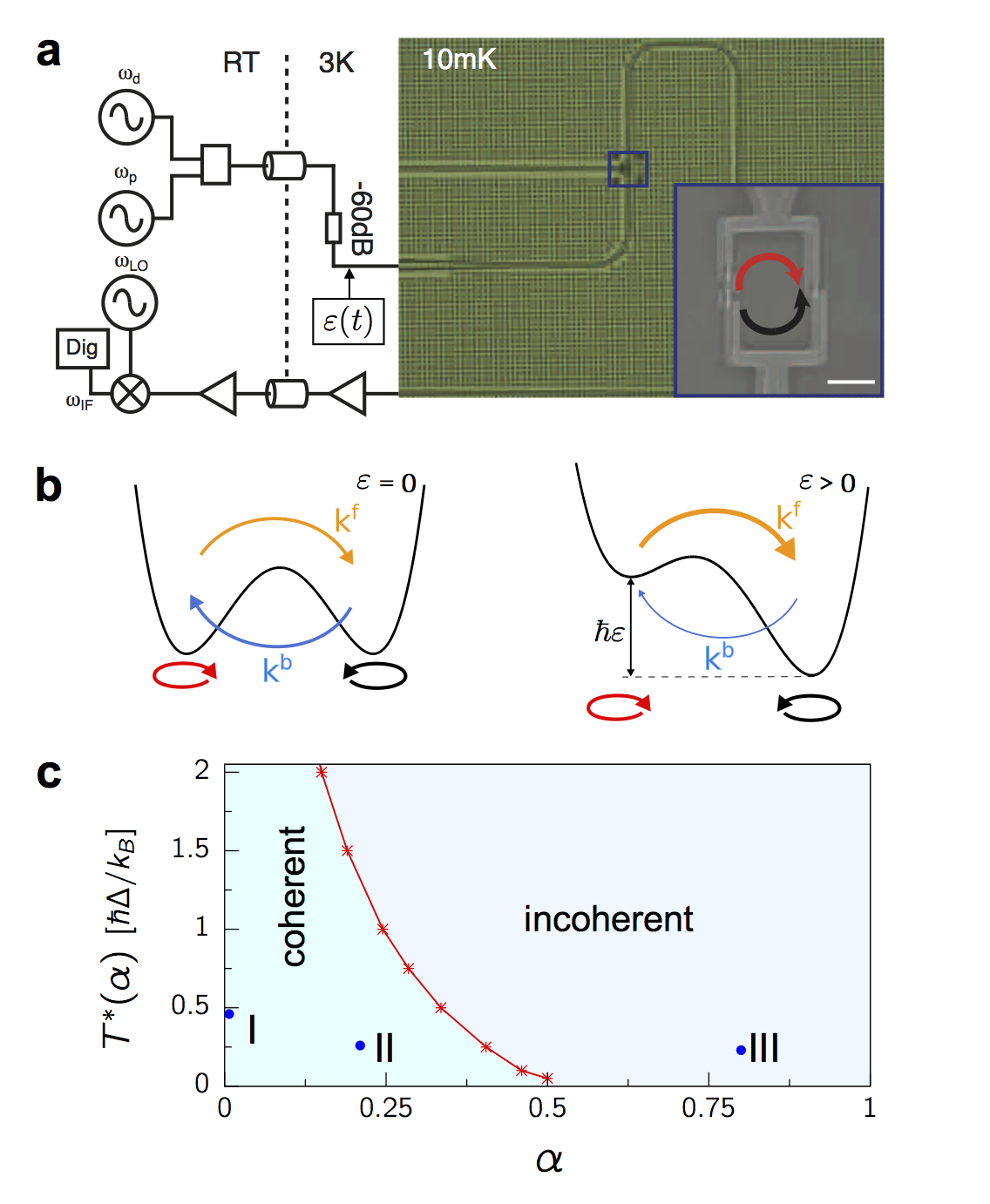}
\caption{\label{fig1} Experimental setup and phase diagram of the symmetric spin-boson model. \textbf{a} Measurement, driving circuit schematic and optical micrograph of a device similar to the ones used in the experiment. A coplanar waveguide running across the chip plays the role of the bath coupled to the qubit. The inset is a scanning electron micrograph showing the qubit attached to the line. The scale bar is 2~$\mu\rm{m}$. Here and in panel \textbf{b} the red (black) arrow indicates clockwise (anticlockwise) circulating persistent currents. 
\textbf{b}, Schematics of the double-well potential associated to the flux threading the qubit. In the absence of external driving sources the potential is symmetric and the forward and backward tunneling rates $k^{\rm f/b}$ are equal. In the presence of a positive bias asymmetry $\varepsilon$, forward tunneling dominates over backward tunneling. 
\textbf{c} Dependence of the temperature $T^*(\alpha)$ for the crossover from the coherent to the incoherent tunneling regime on the coupling $\alpha$. The red curve interpolates numerical results (asterisks) obtained within the nonperturbative NIBA. The dots labeled I, II, and III mark the positions in parameter space of the three devices used in this work. }
\end{figure}

\noindent {\bf Relation between experimental and theoretical observables}\\
\noindent A schematic representation of the experimental setup is shown in Fig.~\ref{fig1}a. The two-state system is a flux qubit, a superconducting circuit consisting of a loop interrupted by four Josephson junctions \cite{Mooij1999}. 
The bosonic environment is formed from electromagnetic modes in the superconducting transmission line coupled to the qubit.
The qubit is pumped by a strong continuous-wave drive  applied through the transmission line. Both the amplitude and the frequency of the drive can be changed over a broad range. The driven system is studied spectroscopically by additionally applying a weak probe field.
 The measured transmission ${\mathcal T}$ at the probe frequency $\omega_{\rm p}$ gives direct access to the linear response function associated to the weak probe signal, the so-called linear susceptibility $\chi$ via the relation
 \begin{equation}
 \label{transmission2}
 {\mathcal T}(\omega_{\rm p})=1-{\rm i}{\cal N}\hbar \omega_{\rm p}\chi(\omega_{\rm p})\;,
 \end{equation}
 where ${\cal N}$ is a coupling constant (see Methods).
 According to Kubo's linear response theory \cite{Kubo1957}, $\chi (\omega)$  carries information about the dispersive and absorptive properties of the qubit in the  absence of the probe, and in turn, as discussed below, about the  dynamical phases of the driven spin-boson system.
By measuring the transmission also when the drive is switched off, we get a reference for the effects of a coherent drive on quantum coherence and localization properties.    \\

\noindent {\bf Phase diagram of the undriven spin-boson model}\\
\noindent We first introduce the spin-boson model and its dynamics without driving. Historically, the Ohmic spin-boson model was first studied in the context of  the tunneling of a quantum particle in a double-well potential  \cite{Leggett1987}. At low temperatures the dynamics are effectively restricted to the Hilbert space spanned by the states $|L\rangle$ and $|R\rangle$, localized in the left and right well, respectively (see Fig.~\ref{fig1}b). Transitions between the two localized configurations are possible due to quantum-mechanical tunneling and are recorded in the time evolution of the population difference $P(t)\equiv \langle\sigma_z(t)\rangle=P_R(t)-P_L(t)$ of the two localized eigenstates. 
 The coordinate associated with the double-well potential need not to be geometrical, but it can represent other continuous variables. For the superconducting flux qubit used in our experiment, this is the magnetic flux  $\Phi$  in the loop. The eigenstates $|L\rangle$ and $|R\rangle$ of the flux operator are related to currents circulating clockwise/anticlockwise  in the superconducting loop~\cite{Mooij1999} (see red/black arrows in Fig.~\ref{fig1}a, b). 
In this basis, the qubit Hamiltonian is 
\begin{eqnarray}\label{Hqb}
H_{\text{qb}}(t)&=&-\frac{\hbar}{2}\left[\Delta \sigma_x+\varepsilon(t)\sigma_z\right]\;,
\end{eqnarray}
where $\sigma_i$ are the Pauli matrices. The parameter $\Delta$ accounts for interwell tunneling and $\hbar\varepsilon(t)$  is the difference in energy between the two wells, which is controllable. The electromagnetic field in the transmission line can be described as a continuously distributed set of propagating modes with a distribution in frequency given by the spectral density 
 \begin{eqnarray}\label{G}
 G(\omega)=2\alpha\omega e^{-\omega/\omega_{\rm c}}\;,
\end{eqnarray}
corresponding to Ohmic damping with the dimensionless coupling strength $\alpha$ and high frequency cutoff $\omega_{\rm c}$.
 
\begin{figure*}[ht!]
\includegraphics[width=16.5cm]{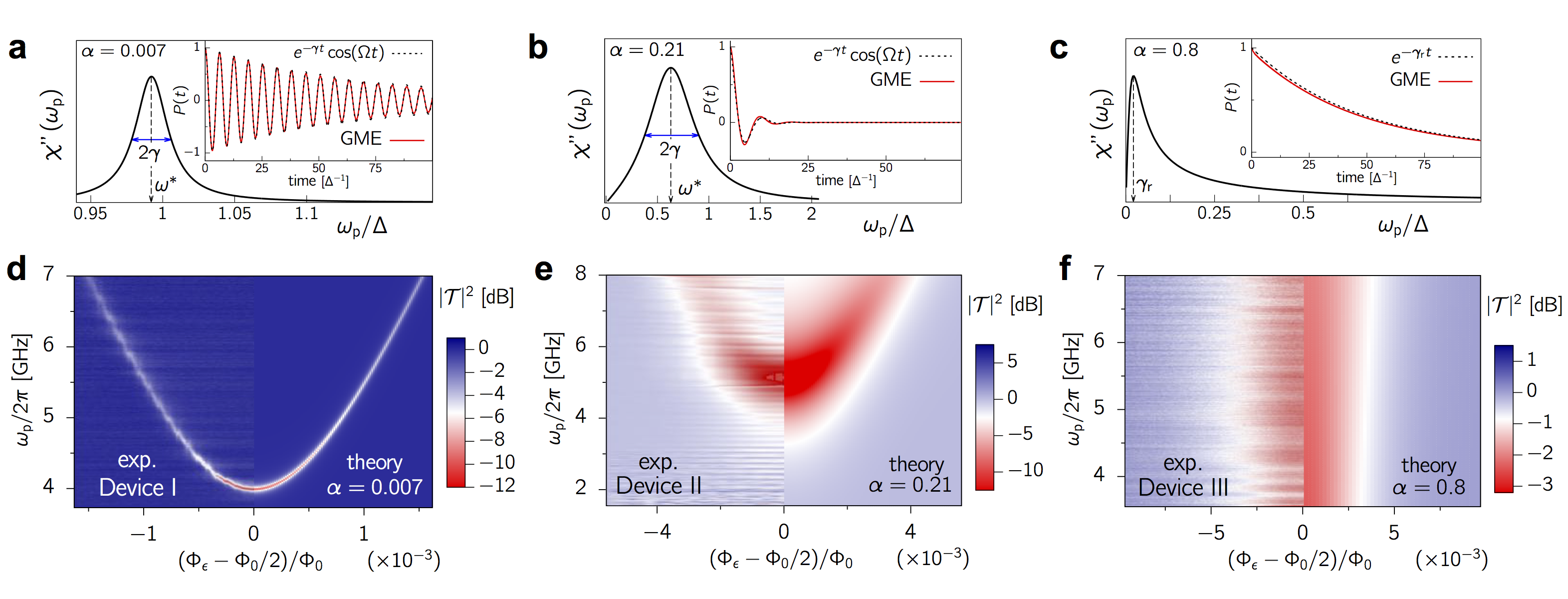}
\caption{\label{fig2} Spin-boson dynamics and spectra at different coupling strengths in the absence of the drive. \textbf{a-c} Frequency dependence of the imaginary part $\chi''(\omega_{\rm p})$ of the linear susceptibility (a. u.) and time evolution of the population difference $P(t)$ (insets) for the three selected combinations of coupling and temperature shown in Fig.~\ref{fig1}c. The position $\omega^*$ and FWHM $2\gamma$ of the linear susceptibility peak in the coherent regimes ($\alpha =0.007$, $\alpha=0.21$) provide a direct measure of the renormalized qubit frequency $\Omega=\sqrt{(\omega^*)^2-\gamma^2}$. In the incoherent regime ($\alpha=0.8$), the peak position yields the relaxation rate $\gamma_{\rm r}$. \textbf{d-f} Experimental  transmission spectra of three flux qubit devices with different coupling junctions are compared with spectra calculated within the NIBA. The characteristic hyperbolic spectrum of the flux qubit is evident in \textbf{d} and recognizable in \textbf{e}. Its disappearance in \textbf{f} indicates the transition to the incoherent regime. At $\Phi_\epsilon=\Phi_0/2$ the spin-boson system is unbiased, which is the situation of panels \textbf{a-c}.}
\end{figure*}

Theoretical work on the spin-boson model has primarily focused on the temporal dynamics of the spin. Quite generally, independent of the initial state of the qubit and the form of the bath spectral density,  energy exchange with the environment is responsible for equilibration of the qubit with the bath on a time scale given by the relaxation rate $\gamma_{\rm r}$. 
 Furthermore, quantum fluctuations and energy exchange yield dephasing with rate $\gamma$. 
In the Ohmic spin-boson model, low frequency environmental modes also lead to a strong renormalization of the bare qubit tunneling splitting $\Delta$. The renormalized qubit frequency $\Omega$ depends  on the bath temperature and coupling strength $\alpha$, and is always reduced with respect to $\Delta$. This leads to three distinct dynamical regimes. Two of them, occurring for $\alpha <1$, are depicted in Fig.~\ref{fig1}c for  the symmetric spin-boson model shown in the left drawing in Fig.~\ref{fig1}b.
The coherent regime corresponds to  $\Omega > \gamma$.  This occurs for $\alpha <1/2$ and a temperature $T<T^*(\alpha)$. In this regime, for a spin initially localized in the right well ($P(0)=1$), the qubit displays damped coherent oscillations  of frequency $\Omega$, specifically, $P(t)=\exp(-\gamma t)\cos(\Omega t)$ [see insets of Fig.~\ref{fig2}a, b].
At the crossover temperature, the renormalized frequency  $\Omega$ vanishes (see Methods and Eq.~(\ref{Tstar}) there). 
The incoherent regime corresponds to $\alpha <1/2$ and $T>T^*(\alpha)$ or $1/2 < \alpha < 1$. The dynamics are characterized by incoherent tunneling transitions with rates $k^{\rm f/b}$ defined in Sec. III of the Methods [see Fig.~\ref{fig2}b]. Correspondingly, we have $P(t)=e^{-\gamma_{\rm r} t}$, where $\gamma_{\rm r}=k^{\rm f}+k^{\rm b}$ [see inset in Fig.~\ref{fig2}c].
 In the third regime, corresponding to $\alpha>1$, localization occurs. Here, the backward and forward rates are renormalized to zero by the low-frequency bath modes. As shown in Fig.~\ref{fig1}c, in the Ohmic spin-boson model, the dynamics becomes fully incoherent above $\alpha=0.5$ for any value of the temperature. As the coupling approaches this value, any perturbative approach in the coupling fails to describe the physics of the system. Consistently with Ref.~\cite{Forn-Diaz2017}, we refer to the coupling regimes $\alpha > 0.5$ as  ultrastrong.\\
\noindent
Primary scope of this work is to understand how the dynamical phase diagram in Fig.~\ref{fig1}c is modified by a periodic modulation of the detuning. This is a formidable task, since the spin-boson problem with time-periodic detuning cannot be solved analytically in the whole parameter space. Exact solutions exist  for the particular value $\alpha=1/2$  \cite{Grifoni1993}. Recently, an analytical solution was suggested for the case of a spin-boson system with time-periodic tunneling amplitude \cite{Restrepo2016}.\\ 

\noindent{\bf Linear susceptibility  of the driven spin-boson model }\\
To carry out our spectroscopic analysis, we describe the bias between the potential wells in our experimental setup by means of the time-dependent function  
\begin{equation}
\label{drive}
\varepsilon(t)=\varepsilon_0 + \varepsilon_{\rm p}\cos(\omega_{\rm p}t) + \varepsilon_{\rm d}\cos(\omega_{\rm d}t)\;.
\end{equation}
Here, the static component $\varepsilon_0$  is related to the externally applied  flux  $\Phi_{\varepsilon}$  by $\varepsilon_0 \propto (\Phi_\varepsilon-\Phi_0/2)$, with $\Phi_0$ the magnetic flux quantum.   
The remaining contributions account for the probe (p), with amplitude $\varepsilon_{\rm p}$ and frequency $\omega_{\rm p}$,  and the  drive (d), with amplitude $\varepsilon_{\rm d}$ and frequency $\omega_{\rm d}$.
 For details, see the Methods. The central quantity in this work is the linear  susceptibility $\chi(\omega_{\rm p})$, which describes the qubit's response at the probe frequency $\omega_{\rm p}$, see Eq. (\ref{transmission2}). The susceptibility measures deviations of the asymptotic population difference, $P^{\rm as}(t)$, from its value $P_{0}$ in the absence of  the weak probe according to \cite{Grifoni1995} 
 \begin{equation}
 P^{\rm as}(t)=P_0 +\hbar\varepsilon_{\rm p}[\chi(\omega_{\rm p})e^{{\rm i}  \omega_{\rm p}t}+\chi(-\omega_{\rm p})e^{-{\rm i} \omega_{\rm p}t}]\;.
 \end{equation}
In this work, the dynamical quantity $P(t)$, and in turn the susceptibility $\chi(\omega_{\rm p})$, have been calculated within the so-called noninteracting-blip approximation (NIBA). This approximation yields a generalized master equation for $P(t)$ with kernels that are nonperturbative in  $\alpha$. It becomes exact at large temperatures and/or coupling strengths~\cite{Weiss2012}.
Under the assumption that $\omega_{\rm d}$ is large compared to the (renormalized) frequency scales of the spin-boson particle, 
closed expressions for the transient evolution of $P(t)$, as well as for the linear susceptibility of the driven spin-boson system, can be obtained (details in the Methods).\\

\noindent{\bf Characterizing the dynamical regimes of the undriven devices}\\
\indent We first demonstrate in Fig.~\ref{fig2}a-c the connection between the imaginary part, $\chi''(\omega_{\rm p})$, of the susceptibility and $P(t)$ for the symmetric spin-boson model in the presence of the probe only ($\varepsilon_0=\varepsilon_{\rm d}=0$). We choose three distinct values of the coupling, namely $\alpha=0.007, 0.21$, situated in the coherent  regime,  and $\alpha=0.8$ in the incoherent regime [see the three dots indicated in Fig. \ref{fig1}c]. 
 In the coherent regime, $\chi''(\omega_{\rm p})$ has a peak at $\omega^*=(\Omega^2 + \gamma^2 )^{1/2}$ with full width at half maximum (FWHM)  given by $2\gamma$.  In the incoherent regime, the peak is located near zero frequency, at the value of the relaxation rate $\gamma_{\rm r}$. According to Eq.~(\ref{transmission2}), a maximum in $\chi''(\omega_{\rm p})$ corresponds to a minimum in the transmission $\mathcal{T}(\omega_{\rm p})$.
By recording the evolution of the transmission as a function of $\omega_{\rm p}$ and of another external parameter, e.g. the static asymmetry $\varepsilon_0$,  various dynamical regimes can be identified.\\
\indent The theoretically calculated transmission  is presented in Fig.~\ref{fig2}d-f  as a function of the applied static bias $\varepsilon_0$ for the three values  of $\alpha$ discussed above. As expected, the qubit dispersion relation can be traced back in the highly coherent and underdamped  regimes corresponding to $\alpha=0.007 $ and $\alpha=0.21$, respectively. In the overdamped regime, with $\alpha=0.8$, the transmission is nearly independent of $\omega_{\rm p}$.  Finally, comparison with the measured transmission for three distinct tunable devices, named I, II, and III in the following, allows us to position the  three devices as shown in the phase diagram in Fig.~\ref{fig1}c. Temperature, cutoff frequency, renormalized splitting $\Omega$, and  conversion factor ${\cal N}$ are estimated from the experiments. Deviations in the choice of these parameters can yield  variations in the estimate of the coupling strength $\alpha$.
The close agreement between the calculated and measured qubit spectra gives a strong evidence that Device III, with an estimated coupling $\alpha=0.8\pm0.1$ (see Sec.~\ref{Suppl7}), is in the nonperturbative ultrastrong coupling regime, buttressing the conclusion of~\cite{Forn-Diaz2017, Diaz-camacho2016}.  In a recent work \cite{Shi2017} a polaron approach, which is equivalent to the NIBA \cite{Weiss2012}, has been used to provide approximate expressions for the response of an undriven qubit coupled to a transmission line.\\
\begin{figure*}[ht]
\centering
\includegraphics[width=16.5cm]{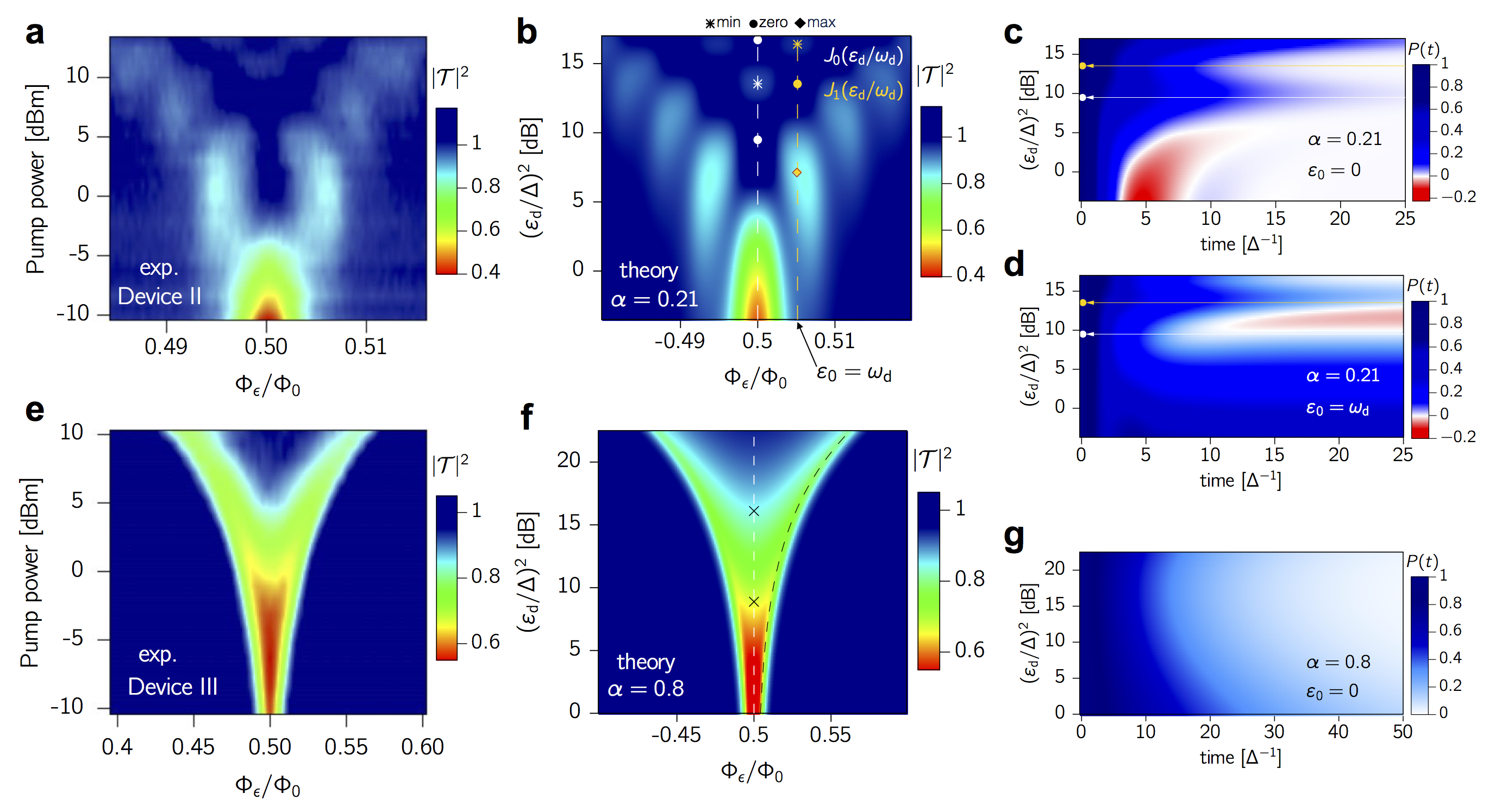}
\caption{\label{fig3} Spectral response and dynamics of the driven spin-boson system. \textbf{a, b} Observed and calculated transmission at the probe frequency for the moderately coupled Device II ($\alpha=0.21$) as function of static bias and pump strength. A clear structure of multi-photon resonances appears. The dashed lines indicate cuts at fixed bias where the dynamics in panels \textbf{c} and \textbf{d} are calculated. \textbf{c, d} Predicted dynamics of $P(t)$ with $P(0)=1$. \textbf{e, f} Observed and calculated spectrum of the ultrastrongly coupled Device III (with $\alpha=0.8$). The spectrum is smoothed, as compared to panels \textbf{a} and \textbf{b},  indicating fully incoherent dynamics. The black dashed line in \textbf{f} corresponds to the condition $\varepsilon_{\rm eff}=2k_{\rm B}T/\hbar$ for the effective nonequilibrium bias [see Eq. (\ref{P0})] and the symbols $\times$ mark the first two zeroes of $J_0(\varepsilon_{\rm d}\tau_{\rm env})$. \textbf{g} Time evolution of $P(t)$ calculated at the symmetry point, $\varepsilon_0=0$, in the same range of pump strengths as in panel \textbf{f}.} 
\end{figure*} 

\noindent{\bf Spectroscopy of the driven spin-boson model}\\
\noindent Let us now turn to the impact of a strong coherent drive on a spin-boson particle in the intermediate and ultrastrong coupling regimes captured by devices II and III, respectively. 
The experimental spectra in Fig.~\ref{fig3}a, e show  the probe transmission as a function of flux bias $\varepsilon_0$ and drive power ($\propto \varepsilon^2_{\rm d}$) for these devices.  Probe and drive frequencies are respectively set to  $\omega_{\rm p}/2\pi = 5.2$ GHz and $\omega_{\rm d}/2\pi=9$ GHz for Device II. For Device III we choose  $\omega_{\rm{p}}/2\pi = 4$~GHz and $\omega_{\rm{d}}/2\pi = 3$~GHz. For Device II, the probe is on-resonance with the undriven qubit at the symmetry point. For Device III, the qualitative features of the driven spectra are largely insensitive to the choice of $\omega_{\rm{p}}$ and $\omega_{\rm{d}}$. The theoretical predictions, shown in Fig.~\ref{fig3}b, f, agree well with the experimental observations. Similar to the pump-only case, striking differences are observed in the transmission of the two devices. 
Let us start discussing Device II. Minima in the transmission are clearly seen in Fig.~\ref{fig3}a, b whenever the static bias matches a multiple of the pump frequency, $\varepsilon_0=n\omega_{\rm d}$, as indicated by the vertical lines drawn in Fig.~\ref{fig3}b for $n=0,1$. Furthermore, the observed pattern with fixed bias at the $n$-th resonance results from a  modulation  by a prefactor proportional to  $J_n(\varepsilon_{\rm d}/\omega_{\rm d})$,  where $J_n$ is a Bessel function of the first kind. For example, the qubit response at the symmetry point is suppressed  in correspondence with the first zero of the Bessel function $ J_0(\varepsilon_{\rm d}/\omega_{\rm d})$  (indicated by a circle), where the incoming probing field is fully transmitted. At larger power, as the zero order Bessel function increases again, the transmission diminishes. Similar patterns have already been reported in driven qubit devices in the highly coherent regime \cite{Wilson2007,Sillanpaa2006}. Those results can be interpreted as a signature of entangled light-matter states known as dressed-states \cite{Nakamura2001,Shevchenko2010,Hausinger2010}.
Near the multiphoton resonance, $\varepsilon_0 = n\omega_{\rm d}$, two of these dressed states form an effective two-level system with dressed tunneling splitting $\Delta_n=\Delta J_n(\varepsilon_{\rm d}/\omega_{\rm d})$. Near a zero of the $n$-th Bessel function, tunneling is strongly suppressed and hence the transmission is maximal. This phenomenon has been dubbed coherent destruction of tunneling in the literature \cite{Grossmann1991}. Dissipation modifies this simple coherent picture, as demonstrated for Device III in Figs.~\ref{fig3}e, f where no Bessel pattern is present and a smooth ``V-shaped'' transmission is observed instead.\\
\section*{\large D\MakeLowercase{iscussion}}
\indent To understand to what extent dissipation modifies the dressed state picture, we have studied the transient dynamics of the population difference $P(t)$ in the presence of  drive only ($\varepsilon_{\rm p}=0$). As discussed in the Methods, $P(t)$ is governed by a generalized master equation featuring the two nonequilibrium kernels $\mathcal{K}^{+/-}(t)$ which, in the absence of probe field, are symmetric/antisymmetric in the static bias $\varepsilon_0$.
In Laplace space, by solving the pole equation $\lambda+ K^+(\lambda)=0$, where  $K^+(\lambda)=\int_0^\infty \exp(-\lambda t)\mathcal{K}^{+}(t)$, the phase diagram of the driven spin-boson particle can in principle be found along the lines discussed in the Methods.
The kernel $K^+(\lambda)$ can be expressed as the sum $K^{\rm f}(\lambda)+K^{\rm b}(\lambda)$  of the nonequilibrium forward and backward kernels 
\begin{equation}
K^{\rm f/b}(\lambda)=\frac{\Delta^2}{2}\int_{0}^\infty dt\; e^{-Q'(t)-\lambda t}J_0\left[d(t) \right]\cos[Q''(t)\mp\varepsilon_0 t]\;,
\label{forward}
\end{equation}
with $d(t)=2\varepsilon_{\rm d}\omega_{\rm d}^{-1}\sin\left(\omega_{\rm d} t / 2\right)$.
The correlation function $Q(t)=Q'(t)+{\rm i}Q''(t)$ describes the environmental influence and its explicit form is discussed in Sec.~\ref{Suppl1} and in Eqs.~(\ref{Qsl1-a})-(\ref{Qsl1-b}) of the Methods. 
For the present discussion, it is enough to observe that in the long-time limit $t \gg \tau_{\rm env}$, where $\tau_{\rm env}=(2\pi\alpha k_{\rm B} T/\hbar)^{-1}$, the real part of $Q(t)$ assumes the form $Q'(t)\sim t/\tau_{\rm env}+{\rm const.}$ appropriate to white noise. Thus, $\tau_{\rm env}$ yields an estimate  of the memory time of the kernels entering Eq.~(\ref{forward}). 
The impact of the  drive is encapsulated in the time-dependent argument of the Bessel function of first kind  $J_0$. Depending on whether $\omega_{\rm d}\tau_{\rm env} \geq 1$ (slow relaxation) or $\omega_{\rm d}\tau_{\rm env}\leq 1 $ (fast relaxation), two distinct regimes corresponding to devices II and III are encountered, respectively.\\
\indent Let us focus on the first case, explored in Fig.~\ref{fig3}a, b. In this regime, one full cycle of the drive field is possible before environmental effects induce a loss of coherence. Thus, we expect that coherent absorption and emission processes from the drive field take place during a cycle. An expansion of the Bessel function in Eq.~(\ref{forward}) in a Fourier series, $J_0[d(t)]=\sum_{n}J^2_n(\varepsilon_{\rm d}/\omega_{\rm d})\exp({\rm i}n\omega_{\rm d})$, shows that  the channel with $n\omega_{\rm d}=\varepsilon_0$ dominates the series \cite{Grifoni1998}, and hence an effective two-level description with renormalized tunneling splitting $\Delta_n$ applies. A solution of the pole equation in this approximation yields a renormalization of the crossover temperature $T^*(\alpha)\to T^*(\alpha) [J_n(\varepsilon_{\rm d}/\omega_{\rm d})]^{1/(1-\alpha ) }$. Because  $J_n<1$, the pump field always  yields a reduction of quantum coherence. Near the zeros of $J_n$, quantum coherence is fully suppressed and an incoherent decay is expected. This behavior is seen in Fig.~\ref{fig3}c, d, where we show the simulated time evolution of $P(t)$ as a function of pump power  at $\varepsilon_0=0$ and  $\varepsilon_0=\omega_{\rm d}$, respectively.
  The color map of $P(t)$ displays coherent oscillations at low to moderate pump amplitudes, where $J_0(\varepsilon_{\rm d}/\omega_{\rm d})$ is still of  order one. However,  a full suppression of quantum coherence occurs near  the first zero of $J_0$, highlighted by a solid white circle. We notice that the  almost complete standstill  predicted  to occur at the zeros of $J_0$ for a dissipation-free, symmetric two-level  particle \cite{Grossmann1991}, is destroyed by  environmental relaxation processes, albeit on a very slow time scale. 
  A similar suppression of coherence, together with a very slow incoherent decay, is observed 
 at the  first resonance, $\varepsilon_0=\omega_{\rm d}$, shown in Fig.~\ref{fig3}d, in correspondence with the first zero of $J_1$.
  Independently of the initial preparation, the steady state population acquires the value 
  $P_0=(K^{\rm f}-K^{\rm b})/(K^{\rm f}+K^{\rm b})$, where $K^{\rm f/b}=K^{\rm f/b}(\lambda=0)$ are the nonequilibrium  backward and forward rates. For the symmetric case  shown in Fig.~\ref{fig3}c, the backward and forward rates are equal and hence $P_0=0$. 
   A genuine nonequilibrium   behavior is observed in Fig. \ref{fig3}d in the region between the first zeros of $J_0$ and  $J_1$, where   the steady state qubit population $P_0<0$, corresponding to a larger population of the left state despite $\varepsilon_0>0$.
This phenomenon originates from the effective  detailed balance relation
  \begin{equation}
  \label{P0}
  K^{\rm f} = K^{\rm b} e^{ \hbar\varepsilon_{\rm eff}/k_{\rm B}T}
  \end{equation} 
    between the nonequilibrium backward and forward rates $K^{\rm f/b}$. This equation implicitly defines   the effective asymmetry $ \varepsilon_{\rm eff}$. 
Only in the absence of the drive does $ \varepsilon_{\rm eff}$ coincide with the static bias $ \varepsilon_0$. We note that the use of an external coherent drive to tune the direction of long-range electron chemical reactions via a drive-induced effective bias was originally proposed in \cite{Dakhnovskii1995,Goychuk1996}.\\
\indent Let us turn to the explanation of the results for Device III displayed in Fig.~\ref{fig3}e-g, where $\omega_{\rm d}\tau_{\rm env}\ll 1$ applies. 
In this regime the approximate result 
\begin{equation}\label{chi}
\chi(\omega_{\rm p})=\frac{1}{4k_{\rm B}T}\frac{\partial \varepsilon_{\rm eff}/\partial \varepsilon_0}{\cosh^2(\hbar\varepsilon_{\rm eff}/2k_{\rm B}T)}\frac{\gamma_{\rm d}}{\gamma_{\rm d}+{\rm i}\omega_{\rm p}}\;
  \end{equation}
can be obtained from the exact expression Eq.~(\ref{chi2}) of the Methods. This form is associated to the incoherent dynamics of the spin boson particle with 
 nonequilibrium relaxation rate $\gamma_{\rm d}\equiv K^{\rm f}+K^{\rm b}$. At the symmetry point we have $\varepsilon_{\rm eff}=\varepsilon_0=0$, with $\lim_{\varepsilon_0\to 0}\partial \varepsilon_{\rm eff}/\partial \varepsilon_0\neq 0$.
 Correspondingly, the susceptibility $\chi''(\omega_{\rm p})$  has a peak at $\omega_{\rm p}=\gamma_{\rm d}$. 
An expansion in the small parameter $\omega_{\rm d}\tau_{\rm env}$ yields $J_0[d(t)]\approx J_0(\varepsilon_{\rm d} t)$ and hence a relaxation rate  $\gamma_{\rm d}$ which is independent of the driving frequency $\omega_{\rm d}$, consistent with the experimental observation that the spectra depend weakly on $\omega_{\rm d}$. The dependence on the pump amplitude $\varepsilon_{\rm d}$ remains, as clearly seen in Fig.~\ref{fig3}e-g where the transmission at the symmetry point smoothly increases for increasing   drive amplitude. The transmission is almost complete for drive powers above the value $(\varepsilon_{\rm d}/\Delta)^2\simeq16$~dB roughly corresponding to the second zero of $J_0(\varepsilon_{\rm d}\tau_{\rm env})$  (see Fig.~\ref{fig3}f, where the black crosses highlight the first two zeroes). Regarding the transmission at finite static bias,
 we expect that no thermally assisted excitation is possible when $\hbar \varepsilon_{\rm eff}\gg k_{\rm B}T$; correspondingly the susceptibility vanishes, as accounted by the term
  $\cosh^{-2}(\hbar \varepsilon_{\rm eff}/2 k_{\rm B}T)$ in Eq.~(\ref{chi}).
   This behavior is clearly seen in Fig.~\ref{fig3}f, where the black dashed line corresponds to the condition $\hbar\varepsilon_{\rm eff}=2 k_{B}T$. Below the dashed line the effective bias  is larger than the temperature and the signal is fully transmitted.\\
\indent In conclusion, we have experimentally and theoretically explored the paradigmatic driven spin-boson model in the underdamped and  ultrastrong  dynamical regimes. 
Quantum coherence is generally reduced or even destroyed by a drive field in a way  which can be tuned by sweeping the drive amplitude and frequency. 
The control of the dynamics is possible for a generic Ohmic spin-boson particle,  independently of its microscopic details. Localization and even population inversion can be attained by properly tuning the parameters of the coherent drive. 
 Our results might find application in various physical, chemical and quantum biology realizations of the driven spin-boson model.
 \\
  
\section*{Methods}

\noindent {\bf Experimental fabrication and measurement setup}\\
Devices were fabricated according to the procedure explained in Ref. \cite{Forn-Diaz2017}. 
Our setup was designed in such a way that the reservoir (the photons in the transmission line) can still be considered in equilibrium despite the strong pumping applied to the qubit. The response of the photons depends on the intensity of the drive  and on the coupling mechanisms. In our experiment, the degrees of freedom of the bath are very weakly coupled to the drive, compared to the qubit. Hence, even though the qubit is strongly driven, the bath is not. To be more quantitative, the most sensitive component of our bath€ is the 50 Ohm input of our amplifier. From its data sheet, the amplifier starts to become nonlinear for an input power of -12 dBm (its 1dB compression point), which is many orders of magnitude higher than what our pump power is. The other components of our bath, which would be microwave attenuators (resistors), are linear up energies  a few orders of magnitude higher.
From the theoretical point of view, we expect that the transmission of the  fully-driven spin-boson model would differ qualitatively from the one of the system-driven spin-boson model considered in this work. No trivial mapping exists between the two models. The very good agreement between theoretical predictions and the experiment validate our conclusion that merely the system is driven.

\noindent {\bf Relation between theoretical and experimental observables} \\ 
The flux operator in the qubit basis is identified  with $\hat{\Phi}=f \sigma_z$. The proportionality constant $f$ is a fitting parameter which, for low couplings, is estimated to be $f=MI_{\rm pers}$, with $M$ the qubit-line mutual inductance and $I_{\rm pers}$ the persistent current in the superconducting loop. This estimate provides values (see Table~1) which are not far from those obtained from fit to data for devices I and II and from qualitative analysis for Device III.
The externally applied tunable flux  $\Phi_{\varepsilon}$ is related to the static bias by $\hbar\varepsilon_0=2I_{\rm pers} (\Phi_\varepsilon-\Phi_0/2)$, with $\Phi_0$ the magnetic flux quantum.   
The probe  input voltage is connected to the angular frequency $\varepsilon_{\rm p}$ yielding the theoretical probe amplitude, see Eq.~(\ref{drive}), through $V_{\rm p}^{\rm in}(t)=f_{\rm Z}\varepsilon_{\rm p}\cos(\omega_{\rm p} t)$, where the proportionality constant is $f_{\rm Z}=\hbar Z/f$ and $Z$ is the line impedance.  It follows that the constant $\mathcal{N}$ in Eq.~(\ref{transmission2}) is given by the ratio $f/f_{\rm Z}$.\\

\noindent {\bf Parameters used in the simulation}\\
The parameters used in the numerical simulations shown in Figs.~\ref{fig2} and~\ref{fig3} are provided in Table~1. Coupling $\alpha$, bare tunneling frequency $\Delta$, and proportionality constant $\mathcal{N}$ are determined by  fit to data of $|\mathcal{T}|^2$ {\it vs}. $\omega_{\rm p}$ performed for the nondriven devices I and II at the symmetry point $\Phi_\epsilon=\Phi_0/2$ [see Fig.~\ref{fig2}d-e]. Such fits along with their accuracy are shown in Fig.~\ref{Sfits}. In Fig.~\ref{fig2}, the measured value of $90$~mK is used for the temperature. Temperature values used in Fig.~\ref{fig3} account for a possibly higher effective temperature introduced by the drive at the qubit position.
Specifically, for Device II, in the presence of the pump drive, a better qualitative agreement between simulated and experimental transmission is obtained by assuming a higher  temperature.
As the qualitative features of the simulated transmission for Device III, operating at ultrastrong coupling, are weakly sensitive to variations of the temperature, we used the same value of temperature for the pump-probe and the probe-only cases.\\

   \begin{table}
   \textbf{Table 1.  Parameters used for simulations}\\
   \vspace{0.3cm}
   \begin{tabular}{ | l | l | l | p{2cm} |}
    \hline
    $\quad$\textbf{Fig.~\ref{fig2}} \&~\textbf{\ref{fig3}}&{Device I}~ & {Device II}~ & {Device III}~  \\ \hline
       $\omega_{\rm c}/2\pi$ (GHz)&65 & 65 & 65  \\ \hline
    $I_{\rm pers}$ (nA)&600 & 280 & 250  \\ \hline
   $\alpha$&$0.007$ [fit] & $0.21$ [fit] & $0.8^*$  \\ \hline  
    $\Delta/2\pi$ (GHz)&$4.04$ [fit] &$7.23$ [fit] &$8.0^*$ \\ \hline
     $\qquad$\textbf{Fig.~\ref{fig2}}&{Device I}~ & {Device II}~ & {Device III}~  \\ \hline
    $T$ (mK) & 90 & 90 & 90 \\ \hline
     $\mathcal{N}$&$0.03$ [fit] &$1.1$ [fit] &$8.0^*$    \\  
 $\qquad$ (estimated)&$(0.02)$&$(0.5)$&$(5-10)$\\    \hline
    $\qquad$\textbf{Fig.~\ref{fig3}}& & {Device II}~ & {Device III}~  \\ \hline
    $T$ (mK) &  & $175^*$ & $90^*$ \\ \hline
      $\mathcal{N}$&  &$1.1^*$  &$16.0^*$   \\ \hline
       $\omega_{\rm p}/2\pi$ (GHz)&  &$5.2$  &$4.0$    \\ \hline
      $\omega_{\rm d}/2\pi$ (GHz)&&$9.0$  &$3.0$  \\ \hline 
         \end{tabular}\\
          \vspace{0.1cm}
        $^*$\footnotesize{value yielding qualitative agreement with the experiment, see Sec.~\ref{Suppl7}.}
    \end{table}

\noindent {\bf Driven spin-boson dynamics within the NIBA}\\
The spin-boson model describes the coupling of a two-level quantum system to a bath of  harmonic oscillators \cite{Caldeira1983}.  
By assuming a coupling  which linearly depends on the  coordinates of the oscillators, one arrives at the famous spin-boson  Hamiltonian 
\begin{equation}\label{H}
H(t)=H_{\text{qb}}(t)-\frac{\hbar}{2}\sigma_z\sum_i c_i (a_i^{\dagger}+a_i)+\sum_i\hbar\omega_i a_i^{\dagger}a_i\;,
\end{equation}
where $a_i$, $a^\dagger_i$ are bosonic annihilation and creation operators and the coefficients $c_i$ are the amplitude of the interaction strength of the two-level system with mode $i$. The bosonic heat bath is fully characterized by the spectral function $ G(\omega)=\sum_i c_i^2\delta(\omega-\omega_i)$. For Ohmic damping,  $G(\omega)\propto \omega$, as assumed in Eq.~(\ref{G}). \\
\indent  The Ohmic spin-boson problem owes its popularity to its ubiquity and to the variety of parameter regimes it encompasses as the temperature $T$ and the coupling strength $\alpha$ are varied.
 We refer the readers to Ref.~\cite{Weiss2012} for an exhaustive treatment.
The dynamical properties of a driven spin-boson system in the strongly damped and in the incoherent regimes, are well described within the so-called noninteracting-blip approximation (NIBA). Furthermore, the NIBA captures well the dynamics of a symmetric $(\varepsilon_0=0)$  spin-boson system in the whole parameter regime. The NIBA approximation provides a generalized master equation (GME) for the evolution of the population difference $P(t)$ with rates in second order in the bare tunneling splitting $\Delta$ but nonperturbative in $\alpha$. Accounting for the presence of time dependent fields, the GME explicitly reads
\begin{eqnarray}\label{GME}
\dot{P}(t)=\int_{t_0}^tdt'\left[ {\cal K}^{-}(t,t')-{\cal K}^{+}(t,t')P(t')\right].
\end{eqnarray}
The NIBA kernels ${\cal K}^\pm$, averaged over a pump period, are given by
\begin{eqnarray}
{\cal K}^{+}(t,t')&=& h^{+}(t-t')\cos\left[\zeta(t,t')\right],\label{K-a}\\
{\cal K}^{-}(t,t')&=& h^{-}(t-t')\sin\left[\zeta(t,t')\right],\label{K-b}
\end{eqnarray}
with
\begin{eqnarray}
h^{+}(t)&=& \Delta^2 e^{-Q'(t)}\cos[Q''(t)]J_0\left[\frac{2\varepsilon_{\rm d}}{\omega_{\rm d}}\sin\left(\frac{\omega_{\rm d} t}{2}\right) \right],\label{hpm-a}\\
h^{-}(t)&=& \Delta^2 e^{-Q'(t)}\sin[Q''(t)]J_0\left[\frac{2\varepsilon_{\rm d}}{\omega_{\rm d}}\sin\left(\frac{\omega_{\rm d} t}{2}\right) \right].\label{hpm-b}
\end{eqnarray}
The function $Q(t)=Q'(t)+{\rm i}Q''(t)$ is the environmental correlation function. For the Ohmic spectral density function $G(\omega)=2\alpha\omega\exp(-\omega/\omega_{\rm c})$,  $\alpha$ being the dimensionless coupling strength and $\omega_{\rm c}$ a high frequency cutoff, and in the scaling limit
$\hbar\omega_{\rm c}\gg \beta^{-1}=k_{\rm B} T$, these functions have an explicit form~\cite{Weiss2012}
\begin{eqnarray}
Q'(t)&=&2\alpha\ln\left[\sqrt{1+\omega_{c}^{2} t^{2}}\frac{\sinh(\pi  t/\hbar\beta)}{\pi t/\hbar\beta}\right],\label{Qsl1-a}\\
Q''(t)&=&2\alpha\arctan(\omega_{c}t).\label{Qsl1-b}
\end{eqnarray}
The above formulas are  accurate in all coupling regimes, provided that the cutoff frequency is large with respect to the other frequency scales involved. In the long-time limit ($t/\beta\hbar\gg1$) the real part of $Q(t)$ assumes the form $Q'(t)\sim t/\tau_{\rm env}+{\rm const.}$, where $\tau_{\rm env}=(2\pi\alpha k_{\rm B} T/\hbar)^{-1}$. Thus the latter quantity determines the memory time of the kernels ${\cal K}^{\pm}$ in Eqs.~(\ref{K-a})-(\ref{K-b}).\\
\indent The dynamical phase entering the kernels reads 
{\begin{eqnarray}\label{zeta}
\zeta(t,t')=(t-t')\varepsilon_0 +\frac{\varepsilon_{\rm p}}{\omega_{\rm p}}\left\{\sin(\omega_{\rm p} t)-\sin\left[\omega_{\rm p} (t')\right]\right\}.
\end{eqnarray}
Note that  in the absence of the probe field, $\varepsilon_{\rm p}=0$, the pump-averaged kernels depend only on the difference $t-t'$,  i.e., ${\cal K}^{\pm}(t,t')={\cal K}^{\pm}(t-t')$, as in the static case. The latter is then recovered by additionally setting $\varepsilon_{\rm d}=0$. On the other hand, the probe-only setup is described by Eq.~(\ref{GME}) upon setting $\varepsilon_{\rm d}=0$ in Eqs.~(\ref{hpm-a})-(\ref{hpm-b}). 
  The dynamics shown in the insets of Fig.~\ref{fig2}a-c are based on the numerical solution of the GME~(\ref{GME}) for $\varepsilon(t)=0$, whereas in the time evolution of $P(t)$ {\it vs}. pump power shown in panels c, d, and g of Fig.~\ref{fig3}, only the probe field is set  to zero. \\

\noindent  {\bf The linear susceptibility}\\
The linear susceptibility is related to the asymptotic probability difference by
\begin{equation}  
\label{p1}
 P^{\rm as}(t)=P_{0}+\hbar \varepsilon_{\rm  p}[\chi(\omega_{\rm p})e^{{\rm i}\omega_{\rm p}t}+\chi(-\omega_{\rm p})e^{-{\rm i}\omega_{\rm p}t}]\;,
 \end{equation}
where, in the NIBA, $P_0$ reduces to the equilibrium value $P_{\rm eq}=\tanh(\hbar\varepsilon_0/2k_{\rm B}T)$ in the absence of pump driving.
The transmission ${\mathcal T}(\omega_{\rm p})$ and the  susceptibility $\chi (\omega_{\rm p})$  shown in the theoretical plots of Figs.~\ref{fig2} and \ref{fig3} are calculated by means of
the exact NIBA  expression \cite{Grifoni1998}  
\begin{equation}\label{chi2}
P_0=\frac{K^-(0)}{K^+(0)}\;, \qquad
 \chi(\omega_{\rm p})
=\frac{H^+(\omega_{\rm p}) - H^-(\omega_{\rm p})P_0}{{\rm i}\omega_{\rm p}+{K}^+({\rm i}\omega_{\rm p})}\;,
\end{equation}
with superscripts $\pm$ denoting symmetric/antisymmetric functions of $\varepsilon_0$. For our pump-probe  case we find 
\begin{eqnarray}
H^+(\omega_{\rm p})&=& \frac{1}{\hbar\omega_{\rm p}}\int_0^{\infty}dt\;e^{-{\rm i}\omega_{\rm p} t/2}\sin\left(\frac{\omega_{\rm p} t}{2}\right)h^{-}(t)\cos(\varepsilon_0 t) \;,\label{kpm-a}\\
H^-(\omega_{\rm p})&=& \frac{-1}{\hbar\omega_{\rm p}}\int_0^{\infty}dt\;e^{-{\rm i}\omega_{\rm p} t/2}\sin\left(\frac{\omega_{\rm p} t}{2}\right)h^{+}(t)\sin(\varepsilon_0 t) \;, \label{kpm-b}\\
K^{+}(\lambda)&=&\int_0^{\infty}dt\; e^{-\lambda t} h^{+}(t)\cos(\varepsilon_0 t) \label{kpm-c}\\
K^{-}(\lambda)&=&\int_0^{\infty}dt\;e^{-\lambda t} h ^{-}(t)\sin(\varepsilon_0 t)\;.\label{kpm-d}
\end{eqnarray}
Here $K^\pm (\lambda) =\int_0^\infty d\tau e^{-\lambda \tau}  {\cal K}^\pm (\tau)$ are the Laplace transforms of the pump-averaged kernels in Eqs.~(\ref{K-a})-(\ref{K-b}) with $\varepsilon_{\rm p}=0$. The kernels $K^\pm (\lambda)$ are related to the forward and backward rates $K^{\rm f/b}(\lambda)$, introduced in Eq.~(\ref{forward}), by $K^\pm =K^{\rm f}\pm K^{\rm b}$. 
 Also, the incoherent rates for the static case are defined as $k^{\rm f/b}=K^{\rm f/b} (\lambda =0, \varepsilon_{\rm d}=0)$.
For devices I and II, in the absence of pump driving, we  analytically evaluated the integrals in Eqs.~(\ref{kpm-a})-(\ref{kpm-d}) and used the resulting expressions in the susceptibility $\chi$, Eq.~(\ref{chi2}), to perform fits to the data. In the limit  $\omega_{\rm p}\tau_{\rm env}\ll1$, Eq.~(\ref{chi2}) simplifies to Eq.~(\ref{chi}) of the main text (see Sec.~\ref{Suppl4}).\\

\noindent  {\bf Coherent-to-incoherent transition}\\
 In the absence of probe driving, $\varepsilon_{\rm p}=0$, the population difference $P(t)$ is conveniently obtained by introducing the Laplace transform $\hat P(\lambda)=\int_0^\infty dt e^{-\lambda t} P(t)$. From Eq.~(\ref{GME}) one finds 
 \begin{equation}
\hat P(\lambda)=\frac{1-K^-(\lambda)/\lambda}{\lambda + K^+(\lambda)}.
 \label{Plambda}
 \end{equation}
 The pole in $\lambda =0$ determines the asymptotic value $P_0=K^-(0)/K^+(0)$ reached at long times. The solution of the equation $\lambda + K^+(\lambda)=0$ yields information on the transient dynamics. In the underdamped regime, complex solutions yield  the renormalized tunneling frequency with associated dephasing rate. In the incoherent regime, the long-time dynamics is ruled by a single exponential decay with relaxation rate $\gamma_{\rm d}\equiv K^+(\lambda=0)$, see Eq.~(\ref{kpm-c}).  \\
\indent Let us focus exemplarily on the undriven spin-boson system at the symmetry point $\varepsilon_0=0$. Then,  an expansion around $\lambda=0$ yields a quadratic equation for the poles of $\hat P(\lambda)$ \cite{Weiss86}. 
 In the coherent regime  the  roots are complex conjugated,  $\lambda_{1,2}=-\gamma \pm {\rm i}\Omega (T)$, while they are real in the incoherent regime (cf. insets in Fig.~\ref{fig2} a-c).  
 The temperature $T^*$ at which the oscillation frequency $\Omega(T)$ vanishes determines the transition between the coherent and incoherent regimes.
 For weak coupling one finds for example  $\Omega=\Delta_{\rm r}(1- \pi \alpha  \hbar \Delta_{\rm r}/k_{\rm B}T) $ with
 \begin{equation}
\label{Delta_r}
\Delta_{\rm r} = \Delta(\Delta/\omega_{\rm c})^{\alpha/(1-\alpha)}g(\alpha)
\end{equation}
and 
$g(\alpha)=[\Gamma (1-2\alpha)\cos(\pi \alpha)]^{1/2(1-\alpha)}$.
  This allows the estimate $T^*(\alpha) \approx \hbar\Delta_{\rm r}(k_{\rm B} \alpha)^{-1}$ when $\alpha \ll 1$. For general $\alpha <1$ it 
  is given by  
 \begin{equation}\label{Tstar}
T^*(\alpha)\approx \frac{\hbar\Delta_{\rm r}}{k_{\rm B}} [\Gamma (\alpha)/\alpha \Gamma (1-\alpha)]^{1/2(1-\alpha)},
\end{equation}
where $\Gamma(x)$ is the Euler Gamma function. This approximate expression matches well the numerically calculated crossover temperature shown in Fig.~\ref{fig1}c. 
The coherent-incoherent transition temperature $T^*(\alpha)$ depicted there is established, for $\alpha<0.5$, by using Eq.~(\ref{chi2}), with numerically evaluated kernels,  whereas the point at $\alpha=0.5$ is individuated by the exact result $k_{\rm B} T^*(\alpha=0.5)/\hbar\Delta=\Delta/2\omega_{\rm c}$~\cite{Weiss2012}. 
Further details are found in Sec.~\ref{Suppl9}.\\

\section*{\large A\MakeLowercase{cknowledgments}} The authors acknowledge  financial support by the Deutsche Forschungsgemeinschaft via  SFB~631, NSERC of Canada, the Canadian Foundation for Innovation, the Ontario Ministry of Research and Innovation, Industry Canada and Canadian Microelectronics Corporation. L.M. gratefully acknowledges financial support by Angelo Della Riccia Foundation and hospitality by Regensburg University during the early stages of the project. P.F.-D. is supported by the Beatriu de Pin\'os fellowship (2016BP00303).  
The authors thank J.~J.~Garc\'ia-Ripoll, B.~Peropadre, and P. H\"anggi for fruitful discussions, and S.~Chang, A.~M.~Vadiraj and C.~Deng for help with device fabrication and with the measurement setups.

\section*{\large A\MakeLowercase{uthor contributions}}
L.M. and  M.G. performed the theoretical analysis, with numerical simulations carried out by L.M.  The experiments were designed and performed by P.F.-D., A.L., and C.M.W.. The devices were fabricated by P.F.-D., J.-L.O., M.A.Y., and M.R.O. contributed to device design and fabrication. R.B. assisted in numerical modeling of the device. The manuscript was mainly written by M.G. with critical comments provided by all authors. The Supplementary Information was mainly written by L.M..

\clearpage

\def\bibsection{\subsection*{References}}

\setcounter{equation}{0}
\setcounter{figure}{0}
\setcounter{table}{0}
\renewcommand{\thesubsection}{S\arabic{subsection}}   
\renewcommand{\theequation}{S\arabic{equation}} 
\renewcommand{\thefigure}{S~\arabic{figure}}
\hyphenation{spec-tros-co-py}

\section*{Supplementary Information}

\subsection{Generalized master equation for the driven spin-boson model}\label{Suppl1}
The spin-boson model describes a two-level system -- the qubit -- interacting with an environment of quantum harmonic oscillators, the so-called heat bath. \\
The total Hamiltonian of the model reads
\begin{equation}\label{SHSM}
H(t)=-\frac{\hbar}{2}\left[\Delta \sigma_x+\varepsilon(t)\sigma_z\right]-\frac{\hbar}{2}\sigma_z\sum_i c_i (a_i^{\dagger}+a_i)+\sum_i\hbar\omega_i a_i^{\dagger}a_i\;,
\end{equation}
where $\sigma_j$ are Pauli spin operators and $a_i^\dag$ and $a_i$ are bosonic creation and annihilation operators, respectively. The angular frequency $\Delta$ is the bare frequency splitting at zero bias.
Within the noninteracting-blip approximation (NIBA), the time evolution of the qubit's population difference $P(t)=\langle \sigma_z(t)\rangle$ is governed by the following generalized master equation (GME)~\cite{Leggett1987,Grifoni1998,Weiss2012}
\begin{eqnarray}\label{SGMESM}
\dot{P}(t)=\int_{t_0}^tdt'\left[ \mathcal{K}^{-}(t,t')-\mathcal{K}^{+}(t,t')P(t')\right]\;.
\end{eqnarray}
In the presence of a time dependent bias described by $\varepsilon(t)=\varepsilon_0 + \varepsilon_{\rm p}\cos(\omega_{\rm p}t) + \varepsilon_{\rm d}\cos(\omega_{\rm d}t)$, where the subscripts "p" and "d" denote probe and drive, respectively, the exact NIBA kernels are 
\begin{eqnarray}
\mathcal{K}_{\rm N}^{+}(t,t')&=& \Delta^2 e^{-Q'(t-t')}\cos[Q''(t-t')]\cos\left[\zeta_{\rm tot}(t,t')\right]\;,\label{SK-a}\\
\mathcal{K}_{\rm N}^{-}(t,t')&=& \Delta^2 e^{-Q'(t-t')}\sin[Q''(t-t')]\sin\left[\zeta_{\rm tot}(t,t')\right]\;,\label{SK-b}
\end{eqnarray}
where the total dynamical phase has the form 
\begin{eqnarray}\label{SzetatotSM}
\zeta_{\rm tot}(t,t')&=&\int_{t'}^{t}dt''\;\varepsilon(t'')\;.
\end{eqnarray}

Averaging over a period $2\pi/\omega_{\rm d}$ yields an effective description of the drive by means of the following NIBA kernels~\cite{Grifoni1998}, which we use for our calculations 
\begin{eqnarray}
\mathcal{K}^{+}(t,t')&=& h^{+}(t-t')\cos\left[\zeta(t,t')\right]\;,\label{SKSM-a}\\
\mathcal{K}^{-}(t,t')&=& h^{-}(t-t')\sin\left[\zeta(t,t')\right]\;,\label{SKSM-b}
\end{eqnarray}
with the functions $h^{\pm}(t)$ reading
\begin{eqnarray}
h^{+}(t)&=& \Delta^2 e^{-Q'(t)}\cos[Q''(t)]J_0\left[\frac{2\varepsilon_{\rm d}}{\omega_{\rm d}}\sin\left(\frac{\omega_{\rm d} t}{2}\right) \right]\;,\label{ShpmSM-a}\\
h^{-}(t)&=& \Delta^2 e^{-Q'(t)}\sin[Q''(t)]J_0\left[\frac{2\varepsilon_{\rm d}}{\omega_{\rm d}}\sin\left(\frac{\omega_{\rm d} t}{2}\right) \right]\;.\label{ShpmSM-b}
\end{eqnarray}
The dynamical phase 
\begin{eqnarray}\label{SzetaSM}
\zeta(t,t')=\varepsilon_0 (t-t')+\frac{\varepsilon_{\rm p}}{\omega_{\rm p}}\left[\sin(\omega_{\rm p} t)-\sin\left(\omega_{\rm p} t' \right)\right]
\end{eqnarray}
entering the averaged NIBA kernels in Eqs.~(\ref{SKSM-a})-(\ref{SKSM-b}) accounts now exclusively for the static bias and the probe field, whereas the drive is taken into account, in an effective description, by the Bessel functions $J_0$ in the functions of $h^{\pm}(t)$.\\
\indent The functions $Q'$ and $Q''$ in Eqs.~(\ref{SK-a})-(\ref{SK-b}) and~(\ref{ShpmSM-a})-(\ref{ShpmSM-b}), are the real and imaginary part of the bath correlation function $Q(t)$, respectively. For  Ohmic spectral density function $G(\omega)=2\alpha\omega\exp(-\omega/\omega_{\rm c})$,  $\alpha$ being the dimensionless coupling strength and $\omega_{\rm c}$ a  cutoff frequency, these two functions have the following explicit expressions~\cite{Weiss2012}
\begin{eqnarray}
Q'(t)&=&\alpha\ln(1+\omega_{\rm c}^2 t^2)+4\alpha\ln\Bigg|\frac{\Gamma(1+\omega_{\beta}/\omega_{\rm c})}{\Gamma(1+\omega_{\beta}/\omega_{\rm c}+{\rm i} \omega_{\beta} t)}\Bigg|\;,\label{SQ-a}\\
Q''(t)&=&2\alpha\arctan(\omega_{\rm c} t)\;,\label{SQ-b}
\end{eqnarray}
where we have introduced the thermal frequency $\omega_{\beta}=(\hbar\beta)^{-1}$ and where $\Gamma(x)$ is the Euler Gamma function. In the limit $\hbar\omega_{\rm c}\gg k_{\rm B} T$ (or $\omega_{\rm c}\gg \omega_{\beta}$), neglecting the ratio $\omega_{\beta}/\omega_{\rm c}$ and using $\Gamma(1+{\rm i}x)\Gamma(1-{\rm i}x)=\pi x/\sinh(\pi x)$, we get the so-called scaling limit  forms 
\begin{eqnarray}
Q'(t)&=&2\alpha\ln\left[\sqrt{1+\omega_{c}^{2} t^{2}}\frac{\sinh(\pi\omega_{\beta} t)}{\pi\omega_{\beta} t}\right]\;,\label{SQsl1-a}\\
Q''(t)&=&2\alpha\arctan(\omega_{c}t)\;.\label{SQsl1-b}
\end{eqnarray}
These expressions are  accurate in every regime, provided that the cutoff frequency is large with respect to the other frequency scales involved. For $\omega_{\rm c} t \gg1$, these functions assume the approximated forms
\begin{eqnarray}
Q'(t)&\simeq&2\alpha\ln\left[\frac{\omega_{\rm c}}{\pi\omega_{\beta}}\sinh(\pi\omega_{\beta} t)\right]\;,\label{SQsl-a}\\
Q''(t)&\simeq&\pi\alpha\text{sgn}(t)\;.\label{SQsl-b}
\end{eqnarray}
Especially at high temperature, $\omega_{\beta}\sim \Delta$, the cutoff operated by the real part $Q'(t)$ in the kernels, becomes of purely exponential form on a short time scale, see Eq.~(\ref{SQlongtime}) below. Now, this means that, at strong coupling, the kernels go to zero on a rather short time, where the short time behavior of $Q''$, neglected in Eq.~(\ref{SQsl-b}), is relevant. Therefore we will use the approximated expressions in Eqs.~(\ref{SQsl-a})-(\ref{SQsl-b}) only for $\alpha<0.5$. \\
\indent An insight into the different behaviors shown by the two driven setups in Fig.~3 of the main text, is provided by considering  the memory time of the kernels $\mathcal{K}^{\pm}$. To this end, consider the long-time limit of $Q(t)$ in Eqs.~(\ref{SQsl-a})-(\ref{SQsl-b}). Specifically, for $\omega_{\beta} t=t k_{\rm B} T/\hbar  \gg 1$, the real part of $Q(t)$ acquires the form
\begin{eqnarray}\label{SQlongtime}
Q'(t)\sim t/\tau_{\rm env}+{\rm const.}, \qquad{\rm where}\qquad\tau_{\rm env}=(2\pi\alpha k_{\rm B} T/\hbar)^{-1}\;. 
\end{eqnarray}
This form implies that, at fixed, finite temperature, $\tau_{\rm env}$ decreases as the coupling $\alpha$ is increased.  Moreover, in the above limit, the bath force operator $F(t)$ of the quantum Langevin equation for the spin-boson model is delta-correlated, as $\langle F(t)F(0)\rangle\propto\frac{d^2}{dt^2} Q(|t|)$,  where the average is taken with respect to the thermal state of the bath (see Ref.~\cite{Weiss2012} for details). As a consequence, on the time scale dictated by the limit~(\ref{SQlongtime}) the bath is a white noise source.

\subsection{Relating the transmission to the qubit's dynamics}\label{Suppl2}
Consider the situation depicted in Fig.~\ref{Sfig:scheme}, in which the probe voltage field $V^{\rm in}_{\rm p}(t)=f_{\rm Z}\varepsilon_{\rm p}\cos(\omega_{\rm p} t)$, coming from the left, is scattered by the qubit placed at the center of the transmission line. The proportionality constant $f_{\rm Z}$ has dimensions of flux whereas $\varepsilon_{\rm p}$ is an angular frequency. The scattering at the qubit position results in the transmitted field to the right, $V_{\rm transm}(t)$, and a reflected field to the left, $V_{\rm refl}(t)$.  The flux difference across the qubit  is $\delta\Phi(t)=\Phi^{\rm L}(t)-\Phi^{\rm R}(t)$, the flux being related to the voltage by $\Phi(t)=\int_{-\infty}^t dt'\;V(t')$.\\
\begin{figure}[ht!]
\begin{center}
\includegraphics[width=0.4\textwidth,angle=0]{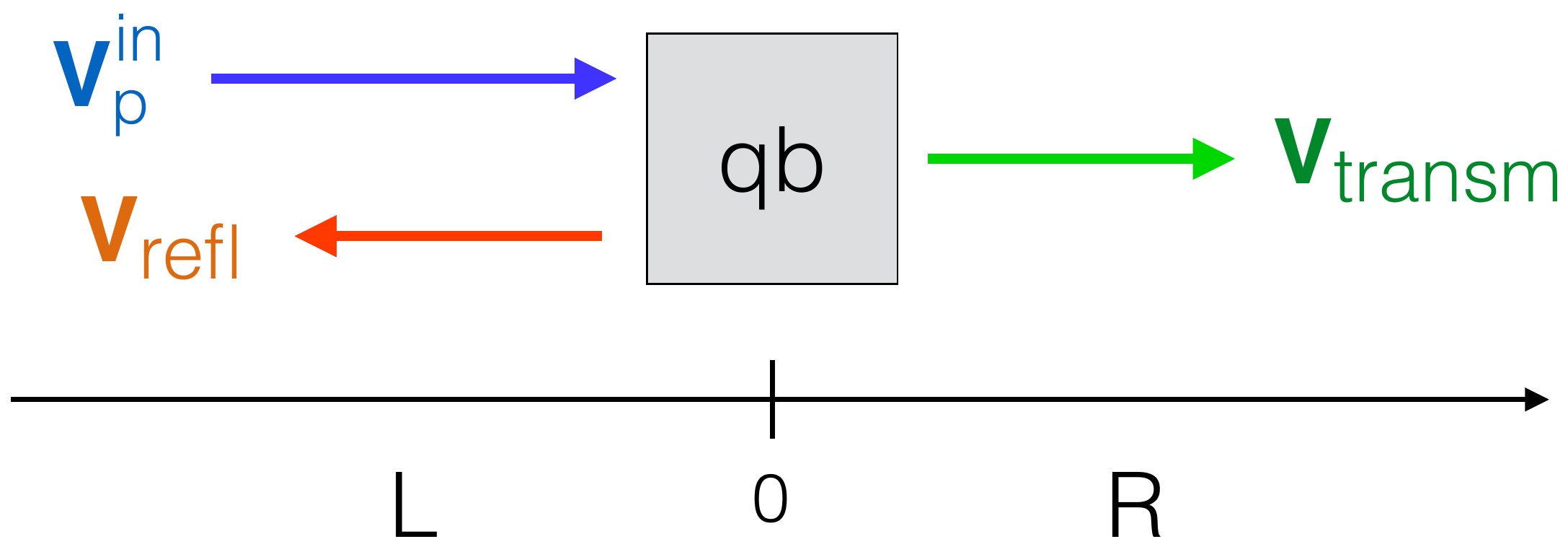}
\caption{\small{Reflection and transmission of the incoming voltage $V^{\rm in}_{\rm p}$.}}
\label{Sfig:scheme}
\end{center}
\end{figure}
\indent A discretized circuit model~\cite{Vool2016} with inductance and capacitance per unit length $l$ and $c$, respectively, gives for the voltage  $V(0^-,t)\equiv V^{\rm L}(t)$ and current $I(0^-,t)\equiv I^{\rm L}(t)$ immediately to the left  of the qubit the following equations
\begin{eqnarray}
V^{\rm L}(t)&=&V^{\rm in}_{\rm p}(t)+V_{\rm refl}(t)\;,\label{SLfields-a}\\
I^{\rm L}(t)&=&\frac{1}{Z}\left[V^{\rm in}_{\rm p}(t)-V_{\rm refl}(t)\right]\;,\label{SLfields-b}
\end{eqnarray}
where $Z=\sqrt{l/c}$ is the characteristic impedance of the transmission line. Similarly, to the right of the qubit, where we set  $V(0^+,t)\equiv V^{\rm R}(t)$ and $I(0^+,t)\equiv I^{\rm R}(t)$, we have
\begin{eqnarray}
V^{\rm R}(t)&=&V_{\rm transm}(t)\;,\label{SRfields-a}\\
I^{\rm R}(t)&=&\frac{1}{Z}V_{\rm transm}(t)\;.\label{SRfields-b}
\end{eqnarray}
Using the conservation of the current, $I^{\rm L}(t)=I^{\rm R}(t)$, and the relation $V^{\rm L}(t)-V^{\rm R}(t)=\dot{\delta\Phi}(t)$, from Eqs.~(\ref{SLfields-a})-(\ref{SRfields-b}) we get
\begin{eqnarray}\label{SVtransm}
V_{\rm transm}(t)&=&V^{\rm in}_{\rm p}(t)-\frac{\dot{\delta\Phi}(t)}{2}\;.
\end{eqnarray}
We identify the flux difference across the qubit with the population difference of the localized eigenstates of the flux operator $\hat{\Phi}=f\sigma_z$, namely we set $\delta\Phi(t)\equiv f\langle\sigma_z(t)\rangle=fP(t)$, where $f$ is the proportionality  constant with dimensions of flux, as described in the main text.

Let $P^{\rm as}(t)=\lim_{t\to\infty}P(t)$ be the asymptotic, nonequilibrium population difference. For periodic driving with period $2\pi/\omega_{\rm p}$, the time derivative $\dot{P}^{\rm as}(t)$ can be expanded as the Fourier series
\begin{eqnarray}\label{SPasdot}
\dot{P}^{\rm as}(t)&=&\sum_m {\rm i}m\omega_{\rm p} p_m e^{{\rm i}m\omega_{\rm p} t}\;,
\end{eqnarray}
where 
\begin{eqnarray}\label{Scoeff}
p_m=\frac{\omega_{\rm p}}{2\pi}\int_{-\mathcal{\pi}/\omega_{\rm p}}^{\mathcal{\pi}/\omega_{\rm p}}dt\; P^{\rm as}(t)e^{-{\rm i}m\omega_{\rm p} t}\;.
\end{eqnarray}
The transmission $\mathcal T$ at frequency $\omega_{\rm p}$ ($m=1$) is defined as  the following ratio between transmitted and input voltages
\begin{eqnarray}\label{St}
\mathcal T(\omega_{\rm p})&=&\frac{V_{\rm transm}(\omega_{\rm p})}{V^{\rm in}_{\rm p}(\omega_{\rm p})}\nonumber\\
&=&\frac{f_{\rm Z}\varepsilon_{\rm p}/2-{\rm i}f\omega_{\rm p} p_1/2}{f_{\rm Z}\varepsilon_{\rm p}/2}\nonumber\\
&=&1-{\rm i}\mathcal{N}\omega_{\rm p}  p_1/\varepsilon_{\rm p}\;,
\end{eqnarray}
where $\mathcal{N}= f/f_{\rm Z}$ and where, in passing from the first to the second line, we used  Eqs.~(\ref{SVtransm}) and (\ref{SPasdot}). Real and imaginary parts of the transmission are therefore given by
\begin{eqnarray}
\text{Re}\{\mathcal T(\omega_{\rm p})\}&=&1+\mathcal{N}\omega_{\rm p} \text{Im}\{p_1\}/\varepsilon_{\rm p}\label{St2-a}\\
\text{and}\qquad\text{Im}\{\mathcal T(\omega_{\rm p})\}&=&-\mathcal{N}\omega_{\rm p} \text{Re}\{p_1\}/\varepsilon_{\rm p}\; ,\label{St2-b}
\end{eqnarray}
respectively.

\subsection{Linear response to a weak probe -- closed expression for the transmission}\label{Suppl3}
In the regime of linear response to an applied monochromatic probe driving, namely for small ratio $\varepsilon_{\rm p}/\omega_{\rm p}$, and within the effective description of the pump drive introduced in Sec.~\ref{Suppl1}, the asymptotic population difference $P^{\rm as}(t)$ is monochromatic~\cite{Grifoni1995,Grifoni1998}. It can be thus expressed as the truncated  Fourier sum
\begin{eqnarray}\label{SPasSM}
P^{\rm as}(t)&\simeq&p_0+p_1^{(1)} e^{{\rm i}\omega_{\rm p} t}+p_{-1}^{(1)} e^{-{\rm i}\omega_{\rm p} t}\nonumber\\
&=&P_0+\hbar \varepsilon_{\rm  pr }[\chi(\omega_{\rm p})e^{{\rm i}\omega_{\rm p}t}+\chi(-\omega_{\rm p})e^{-{\rm i}\omega_{\rm p}t}]\;,
\end{eqnarray}
where the superscript $^{(1)}$ denotes first order with respect to the ratio $\varepsilon_{\rm p}/\omega_{\rm p}$. Here $\chi$ is the linear susceptibility~\cite{Grifoni1995} and $P_0$ is the asymptotic value of $P(t)$ in absence of probe driving. As shown in Fig.~\ref{Sfig:comparison_Pas} below, this constitutes an excellent approximation of the actual dynamics under weak probe driving.  From Eqs.~(\ref{St}) and~(\ref{SPasSM}), the transmission at probe frequency in linear response is related to the dynamical susceptibility by 
\begin{eqnarray}\label{St-linresp}
\mathcal T(\omega_{\rm p})=1-{\rm i}\mathcal{N}\hbar\omega_{\rm p}  \chi(\omega_{\rm p})\;.
\end{eqnarray}
\indent Within the NIBA, by substituting the expression~(\ref{SPasSM}) for $P^{\rm as}(t)$ in the GME~(\ref{SGMESM}), setting the upper integration limit to $t\rightarrow \infty$, which is valid for times much larger than the kernels' memory time, and expanding the kernels in Fourier series, we get the following closed, linear response expression for $p_1^{(1)}$ ~\cite{Grifoni1995,Grifoni1998}
\begin{eqnarray}\label{Sp1SM}
p_1^{(1)}(\omega_{\rm p})&=&\frac{1}{{\rm i}\omega_{\rm p}+v^{+(0)}(\omega_{\rm p})}\left[ k^{-(1)}_1(\omega_{\rm p})-k^{+(1)}_{1}(\omega_{\rm p})\frac{k^{-(0)}_0}{k^{+(0)}_0}\right]
\end{eqnarray}
(superscripts $^{(0,1)}$ denote the order in $\varepsilon_{\rm p}/\omega_{\rm p}$).\\
\indent The kernels $k_m^{\pm}$ and $v^+$, whose approximate forms (perturbative in $\varepsilon_{\rm p}/\omega_{\rm p}$) enter  Eq.~(\ref{Sp1SM}), are defined by 
\begin{eqnarray}
k_m^{\pm}(\omega_{\rm p})&=&\frac{\omega_{\rm p}}{2\pi}\int_{-\pi/\omega_{\rm p}}^{\pi/\omega_{\rm p}}dt\;e^{-{\rm i}m\omega_{\rm p} t}\int_0^{\infty} d\tau\; \mathcal{K}^{\pm}(t,t-\tau)\;,\label{Skm-a}\\
v^{+}(\omega_{\rm p})&=&\frac{\omega_{\rm p}}{2\pi}\int_{-\pi/\omega_{\rm p}}^{\pi/\omega_{\rm p}}dt\;\int_0^{\infty} d\tau\; e^{-{\rm i}\omega_{\rm p} \tau}\mathcal{K}^{\pm}(t,t-\tau)\;,\label{Skm-b}\\
\end{eqnarray}
where the pump drive-averaged kernels $\mathcal{K}^{\pm}(t,t')$ have been introduced in Eqs.~(\ref{SKSM-a})-(\ref{SKSM-b}). 
Expansion of the Bessel functions entering the kernels $\mathcal{K}^{\pm}(t,t')$ to lowest order in $\varepsilon_{\rm p}/\omega_{\rm p}$ by means of  $J_n(x)\sim (x/2)^n$, yields the following explicit expressions for the kernels in Eq.~(\ref{Sp1SM})
\begin{eqnarray}
k^{+(0)}_{0}&=&\int_0^{\infty}dt\; h^{+}(t)\cos(\varepsilon_0 t)\;,\label{SkpmSM-a} \\
k^{-(0)}_{0}&=&\int_0^{\infty}dt\;h^{-}(t)\sin(\varepsilon_0 t)\;,\label{SkpmSM-b} \\
k^{+(1)}_{1}(\omega_{\rm p})&=&-\frac{\varepsilon_{\rm p}}{\omega_{\rm p}}\int_0^{\infty}dt\;e^{-{\rm i}\omega_{\rm p} t/2}h^{+}(t)\sin(\varepsilon_0 t)\sin(\omega_{\rm p} t/2)\;,\label{SkpmSM-c}\\
k^{-(1)}_{1}(\omega_{\rm p})&=&\frac{\varepsilon_{\rm p}}{\omega_{\rm p}}\int_0^{\infty}dt\;e^{-{\rm i}\omega_{\rm p} t/2}h^{-}(t)\cos(\varepsilon_0 t)\sin(\omega_{\rm p} t/2)\;,\label{SkpmSM-d} \\
{\rm and}\quad v^{+(0)}(\omega_{\rm p})&=&\int_0^{\infty}dt\;e^{-{\rm i}\omega_{\rm p} t}h^{+}(t)\cos(\varepsilon_0 t)\;,\label{SkpmSM-e}
\end{eqnarray}
with $h^{\pm}(t)$ defined in Eqs.~(\ref{ShpmSM-a})-(\ref{ShpmSM-b}).  
In Fig.~\ref{Sfig:comparison_Pas} the transient dynamics obtained  from direct integration of the GME~(\ref{SGMESM}) is compared to the asymptotic time-periodic evolution given by Eqs.~(\ref{SPasSM}),~(\ref{Sp1SM}), and~(\ref{SkpmSM-a})-(\ref{SkpmSM-e}).\\
\begin{figure}[ht!]
\begin{center}
\includegraphics[width=0.55\textwidth,angle=0]{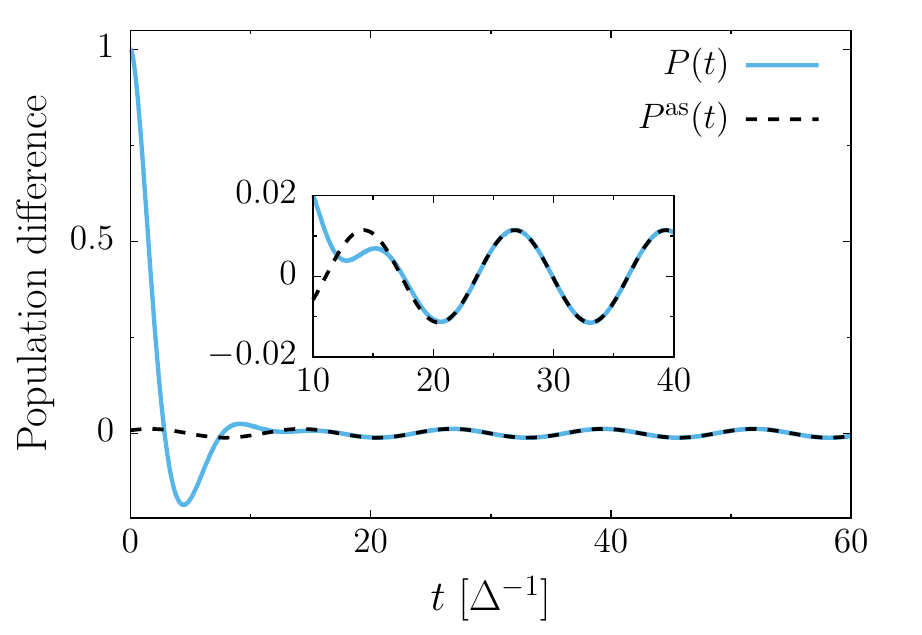}
\caption{\small{Linear response to a weak probe field -- dynamics of the undriven qubit. Time evolution of the population difference $P(t)$ obtained by integrating the GME~(\ref{SGMESM}) with $P(0)=1$ (solid line) compared with the asymptotic dynamics  $P^{\rm as}(t)$ given by Eq.~(\ref{SPasSM}) with $p_1^{(1)}$ from Eq.~(\ref{Sp1SM}) (dashed line). The kernels $k_m^\pm$ and $v^+$ are obtained by numerically evaluating the integrals in Eqs.~(\ref{SkpmSM-a})-(\ref{SkpmSM-e}). The bath correlation function $Q(t)$ in exact scaling limit form  [Eqs.~(\ref{SQsl1-a})-(\ref{SQsl1-b})] is used for both curves. Parameters are $\alpha=0.2$, $T=0.5~\hbar\Delta/k_{\rm B}$, $\omega_{\rm c}=10~\Delta$, $\varepsilon_0=0$, $\varepsilon_{\rm d}=0$, $\omega_{\rm p}=0.5~\Delta$, and $\varepsilon_{\rm p}=0.01~\Delta$.}}
\label{Sfig:comparison_Pas}
\end{center}
\end{figure}
\indent The linear susceptibility $\chi$ is related to the coefficient $p_1^{(1)}$ by Eq.~(\ref{SPasSM}). Thus, from Eq.~(\ref{Sp1SM}), by simplifying the notation, we get 
\begin{equation}\label{SchiSM}
 \chi(\omega_{\rm p})=\frac{H^+(\omega_{\rm p}) - H^-(\omega_{\rm p})P_0}{{\rm i}\omega_{\rm p}+K^+({\rm i}\omega_{\rm p})}\;,\qquad{\rm where} \qquad P_0=K^-(0)/K^+(0)\;.
\end{equation}
Here $K^\pm (\lambda) =\int_0^\infty d\tau e^{-\lambda \tau}  \mathcal{K}^\pm (\tau)$ is the Laplace transform of the pump-averaged kernels with $\varepsilon_{\rm p}=0$. The kernels in Eq.~(\ref{SchiSM}) are related to the ones defined in Eqs.~(\ref{SkpmSM-a})-(\ref{SkpmSM-e}) by   
\begin{eqnarray}\label{Skernels}
K^\pm(\lambda=0)= k^{\pm(0)}_{0}\;,\quad
K^+(\lambda={\rm i}\omega_{\rm p})=v^{+(0)}(\omega_{\rm p})\;,\quad
{\rm and}\quad H^{\pm}(\omega_{\rm p})=\frac{k^{\mp(1)}_{1}(\omega_{\rm p})}{\hbar\varepsilon_{\rm p}}\;.
\end{eqnarray}
Note that, within the present linear response treatment, the transmission is independent of the probe amplitude $\varepsilon_{\rm p}$, cf. Eq.~(\ref{St-linresp}). Note also that the notation for the kernels $H^{\pm}$ reflects the same symmetry with respect to the static bias $\varepsilon_0$ which holds for $K^{\pm}$. Finally, the forward/backward rates
\begin{eqnarray}\label{SforwardSM}
K^{\rm f/b}&=&[K^+(0)\pm K^+(0)]/2\nonumber\\
&=&\frac{\Delta^2}{2}\int_{0}^\infty dt\; e^{-Q'(t)}J_0\left[\frac{2\varepsilon_{\rm d}}{\omega_{\rm d}}\sin\left(\frac{\omega_{\rm d} t}{2}\right) \right]\cos[Q''(t)\mp\varepsilon_0 t]\;,
\end{eqnarray}
introduced in the main text, describe the incoherent tunneling between the individual localized (flux) states.

\subsection{Approximate form of the susceptibility}\label{Suppl4}
 Whenever the condition $\omega_{\rm p}\tau_{\rm env}\ll 1$ is fulfilled, it is possible to expand the kernels $K^+({\rm i}\omega_{\rm p})$ and $H^{\pm}(\omega_{\rm p})$  [see Eq.~(\ref{Skernels})] with respect to $\omega_{\rm p}\tau_{\rm env}$. To first order 
\begin{eqnarray}\label{SHapprox}
K^+({\rm i}\omega_{\rm p})\simeq K^+(\lambda=0)\qquad{\rm and}\qquad H^{\pm}(\omega_{\rm p})\simeq \frac{1}{2\hbar}\frac{\partial}{\partial\varepsilon_0}K^{\mp}(\lambda=0)\;.
\end{eqnarray}
Now, the NIBA prediction for the stationary probability difference $P_0$ in the absence of the pump driving is $P_0=\tanh(\hbar\varepsilon_0/2k_{\rm B}T)$~\cite{Grifoni1998,Weiss2012}.  In the presence of the pump driving, within the present effective description of the pump drive (see Sec.~\ref{Suppl1}), the expression for $P_0$ is generalized as follows 
\begin{eqnarray}\label{SP0}
P_0=\tanh\left(\frac{\hbar\varepsilon_{\rm eff}}{2k_{\rm B}T}\right) \qquad\text{where}\qquad \varepsilon_{\rm eff}=\frac{k_{\rm B}T}{\hbar}\ln \left(\frac{K^{\rm f}}{K^{\rm b}}\right)\;.
\end{eqnarray}
The effective bias $\varepsilon_{\rm eff}$ depends on the static bias $\varepsilon_0$. As a result, in the limit $\omega_{\rm p}\tau_{\rm env}\ll 1$, by substituting the expressions in Eq.~(\ref{SHapprox}) into Eq.~(\ref{SchiSM}) we obtain
\begin{eqnarray}\label{SchiapproxSM}
 \chi(\omega_{\rm p})&\simeq&\frac{K^+(0)}{2\hbar[{\rm i}\omega_{\rm p}+K^+(0)]}\frac{\partial}{\partial\varepsilon_0}\tanh\left(\frac{\hbar\varepsilon_{\rm eff}}{2k_{\rm B}T}\right)\nonumber\\
 &=&\frac{1}{4 k_{\rm B} T}\frac{\partial\varepsilon_{\rm eff}/\partial\varepsilon_0}{\cosh^2(\hbar\varepsilon_{\rm eff}/2k_{\rm B}T)}\frac{\gamma_{\rm d}}{\gamma_{\rm d}+{\rm i}\omega_{\rm p}}\;,
\end{eqnarray}
where $\gamma_{\rm d}=K^+(0)=K^{\rm f}+K^{\rm b}$ (cf. Eq.~\ref{SforwardSM}), and where
\begin{eqnarray}\label{Sderivative}
\frac{\partial\varepsilon_{\rm eff}}{\partial\varepsilon_0}=\frac{k_{\rm B}T}{\hbar}\left(\frac{1}{K^{\rm f}}\frac{\partial K^{\rm f}}{\partial\varepsilon_0}-\frac{1}{K^{\rm b}}\frac{\partial K^{\rm b}}{\partial\varepsilon_0}\right)\;.
\end{eqnarray}
At the symmetry point $\varepsilon_{\rm eff}=\varepsilon_0=0$ so that from Eqs.~(\ref{SforwardSM}) and~(\ref{Sderivative}) we get  
\begin{eqnarray}
\lim_{\varepsilon_0\to 0}\frac{\partial\varepsilon_{\rm eff}}{\partial\varepsilon_0}=\frac{2k_{\rm B}T}{\hbar}\frac{\int_{0}^\infty dt\; t\; h^-(t)}{\int_{0}^\infty dt\; h^+(t)}\;,
\end{eqnarray}
where the functions $h^{\pm}(t)$ have been defined in Eqs.~(\ref{ShpmSM-a})-(\ref{ShpmSM-b}).

\subsection{Analytical evaluation of the kernels in the absence of pump driving}\label{Suppl5}
\indent The integrals in Eqs.~(\ref{SkpmSM-a})-(\ref{SkpmSM-e}) can be solved analytically by using the  bath correlation function $Q(t)$ in the approximated scaling limit form given by Eqs.~(\ref{SQsl-a})-(\ref{SQsl-b}) for $\alpha<1/2$. With this approximated correlation function, the functions $h^{\pm}(t)$, introduced in Eqs.~(\ref{ShpmSM-a})-(\ref{ShpmSM-b}), take on the form
\begin{eqnarray}
h^{+}(t)&=& \Delta^2(2\kappa\omega_{\rm c})^{-2\alpha}\left[\sinh( t/2\kappa)\right]^{-2\alpha}\cos(\pi\alpha)\;,\label{Shpm-approx-a}\\
h^{-}(t)&=&\Delta^2(2\kappa\omega_{\rm c})^{-2\alpha}\left[\sinh( t/2\kappa)\right]^{-2\alpha}\sin(\pi\alpha)\;,\label{Shpm-approx-b}
\end{eqnarray}
where we introduced the time scale $\kappa=\hbar\beta/2\pi=(2\pi\omega_{\beta})^{-1}$.\\
\begin{figure}[ht!]
\begin{center}
\includegraphics[width=0.55\textwidth,angle=0]{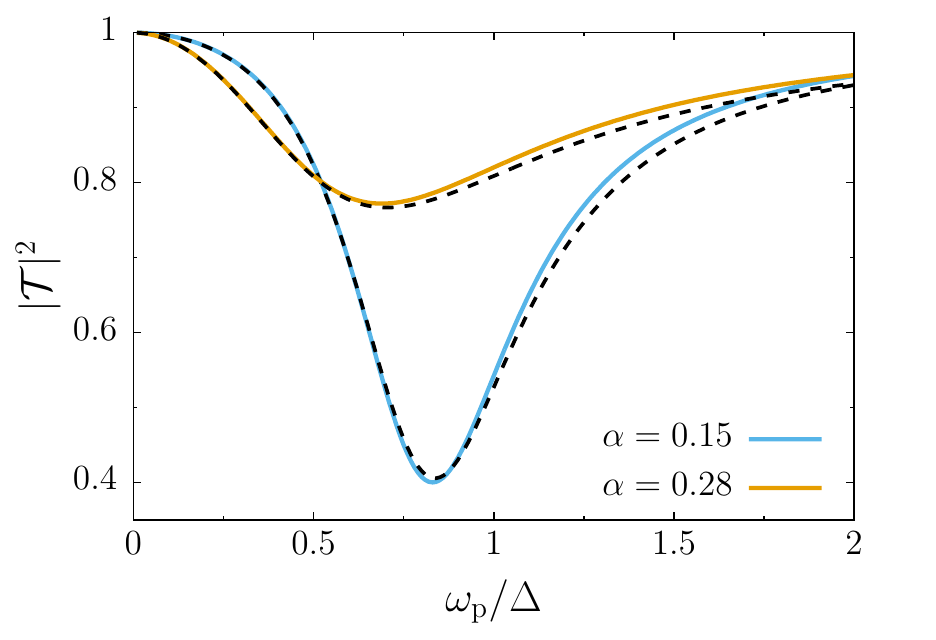}
\caption{\small{Transmission \emph{vs.} probe frequency for the undriven qubit for two values of $\alpha$. The transmission is calculated via Eqs.~(\ref{St-linresp}) and~(\ref{SchiSM}). Solid lines -- kernels numerically evaluated from Eqs.~(\ref{SkpmSM-a})-(\ref{SkpmSM-e}) with the bath correlation function $Q(t)$ in the exact scaling limit from [Eqs.~(\ref{SQsl1-a})-(\ref{SQsl1-b})]. Dashed lines -- kernels in analytical approximated forms [Eqs.~(\ref{SkapproxSM-a})-(\ref{SkapproxSM-d})]. Parameters are $T=0.5~\hbar\Delta/k_{\rm B}$, $\omega_{\rm c}=10~\Delta$, $\varepsilon_0=0$, $\varepsilon_{\rm d}=0$, and $\varepsilon_{\rm p}=0.01~\Delta$.}}
\label{Sfig:comparison_t_mod2}
\end{center}
\end{figure}

\indent We use the exact  result~\cite{Gradshteyn2007}
\begin{equation}
\int_0^{\infty}dt\; e^{-\mu t} \sinh^{\nu}(\beta t)=\frac{1}{2^{\nu+1}\beta}B\left( \frac{\mu}{2\beta}-\frac{\nu}{2},\nu+1\right),
\end{equation}
where $B(x,y)$ is the beta function with the property
\begin{equation}
B(x,y)=\frac{\Gamma(x)\Gamma(y)}{\Gamma(x+y)},
\end{equation}
and $\Gamma(z)$ is the Euler Gamma function, with the property $\Gamma(1-z)\Gamma(z)=\pi/\sin(\pi z)$. By setting $\mu={\rm i}(\omega_{\rm p}\pm\varepsilon_0)$, $\nu=-2\alpha$, and $\beta=(2\kappa)^{-1}$, we obtain the following analytical expression for the kernels in Eq.~(\ref{Skernels})
\begin{eqnarray}
K^{+}(\lambda)&=&N_+\left[ \mathcal{W}(-{\rm i}\lambda+\varepsilon_0)+\mathcal{W}(-{\rm i}\lambda-\varepsilon_0)\right]\;,\label{SkapproxSM-a}\\
K^{-}(0)&=&{\rm i}N_-\left[ \mathcal{W}(\varepsilon_0)-\mathcal{W}(-\varepsilon_0)\right]\;,\label{SkapproxSM-b}\\
H^-(\omega_{\rm p})&=&\frac{1}{2\hbar\omega_{\rm p}}N_+\left[ \mathcal{W}(\omega_{\rm p}+\varepsilon_0)-\mathcal{W}(\omega_{\rm p}-\varepsilon_0)-\mathcal{W}(\varepsilon_0)+\mathcal{W}(-\varepsilon_0)\right]\;,\label{SkapproxSM-c}\\
{\rm and}\quad H^+(\omega_{\rm p})&=&{\rm i}\frac{\varepsilon_{\rm p}}{2\omega_{\rm p}}N_-\left[ \mathcal{W}(\omega_{\rm p}+\varepsilon_0)+\mathcal{W}(\omega_{\rm p}-\varepsilon_0)-\mathcal{W}(\varepsilon_0)-\mathcal{W}(-\varepsilon_0)\right]\;,\label{SkapproxSM-d}
\end{eqnarray}
where
\begin{eqnarray}
N_+&=&\frac{\Delta^2}{2}\frac{\kappa^{1-2\alpha}}{\omega_{\rm c}^{2\alpha}}\cos(\pi\alpha)\Gamma(1-2\alpha)\;,\label{Sfunctions-a}\\
N_-&=&\frac{\Delta^2}{2}\frac{\kappa^{1-2\alpha}}{\omega_{\rm c}^{2\alpha}}\sin(\pi\alpha)\Gamma(1-2\alpha)\;,\label{Sfunctions-b}\\
{\rm and}\quad \mathcal{W}(x)&=&\frac{\Gamma(\alpha+{\rm i}\kappa x)}{\Gamma(1-\alpha+{\rm i}\kappa x)}\;.\label{Sfunctions-c}
\end{eqnarray}
Note that $\mathcal{W}(-x)=\mathcal{W}^*(x)$. 
In Fig.~\ref{Sfig:comparison_t_mod2}, the transmission obtained by using the analytical expressions in Eqs.~(\ref{SkapproxSM-a})-(\ref{SkapproxSM-d}) is compared with the corresponding numerical evaluations of Eqs.~(\ref{SkpmSM-a})-(\ref{SkpmSM-e}) with $Q(t)$ in the exact scaling limit form [Eqs.~(\ref{SQsl1-a})-(\ref{SQsl1-b})].\\
\indent The analytical expressions in Eqs.~(\ref{SkapproxSM-a})-(\ref{SkapproxSM-d}) are used to perform fits to data (see the next section) and for the theory colormaps for devices I and II in Fig.~2  of the main text ($\alpha=0.007$ and $0.21$, respectively, and $\varepsilon_{\rm d}=0$).  This is not the case for Device III in the same figure ($\alpha=0.8$, $\varepsilon_{\rm d}=0$) and for the theory panels in Fig.~3 ($\varepsilon_{\rm d}\neq0$) of the main text, where the transmission is calculated by numerically evaluating the integrals in Eqs.~(\ref{SkpmSM-a})-(\ref{SkpmSM-e}) with correlation function $Q(t)$ in the exact scaling limit form. 

\clearpage

\subsection{Fit to data for devices I and II in the absence of pump driving}\label{Suppl6}
\indent In Fig.~\ref{Sfits} we show the results of fits to the measured transmission at the symmetry point ($\Phi_{\epsilon}=\Phi_0/2$, where $\Phi_{\epsilon}$ is the control field associated with the static bias) for the devices I and II. The spectra of these devices in the absence of drive are shown in Fig.~2 of the main text. Fits to data are performed by using Eqs.~(\ref{St-linresp}) and~(\ref{SchiSM}) with the analytical expressions in Eqs.~(\ref{SkapproxSM-a})-(\ref{SkapproxSM-d}) for the kernels $K^{\pm}$ and $H^{\pm}$. Note that the present treatment has as input the bare value of $\Delta$, the qubit splitting at zero bias, which is not accessed directly in experiments. For this reason $\Delta$ is left as a free parameter, along with the spin-boson coupling $\alpha$ and the prefactor $\mathcal{N}$ in Eq.~(\ref{St-linresp}). Temperature  and cutoff frequency are fixed to $T=90$ mK and $\omega_{\rm c}/2\pi= 65$ GHz, respectively. \\
\begin{figure}[ht!]
\begin{center}
\includegraphics[width=0.95\textwidth,angle=0]{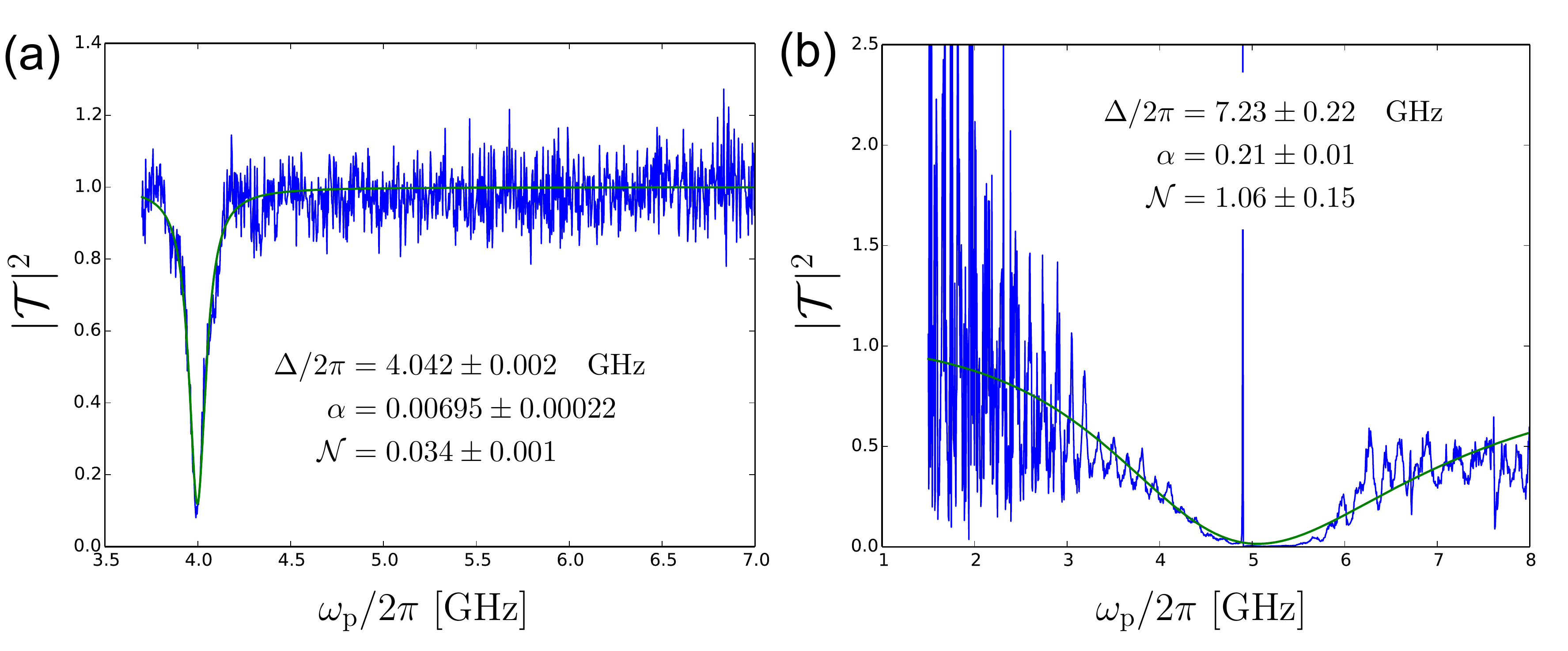}
\caption{\small{Transmission $| \mathcal T|^2$ \emph{vs.} probe frequency at the symmetry point ($\Phi_{\epsilon}=\Phi_0/2$).  The results of fit to the transmission data from experiments provide estimates for $\alpha$, $\Delta$, and $\mathcal{N}$. The analytical expressions in Eqs.~(\ref{SkapproxSM-a})-(\ref{SkapproxSM-d}) are used.  (a) -- Cut at the symmetry point of the spectrum of Device I (see Fig.~2d of the main text). (b) --  Cut at the symmetry point of the spectrum of Device II (see Fig.~2e of the main text).  In both panels, the (fixed)  temperature  and cutoff frequency are $T=90$ mK and $\omega_{\rm c}/2\pi= 65$ GHz, respectively.}}
\label{Sfits}
\end{center}
\end{figure}

\subsection{Estimates for the parameters of Device III}\label{Suppl7}
Device III is in a coupling regime that does not allow for an analytical evaluation of the kernels entering the expression for the transmission [see Eqs.~(\ref{St-linresp})-(\ref{Skernels})]. As a consequence, we are not able to extract via fit to data the parameters that characterize the coupling regime of Device III, as done for devices I and II. Moreover, the spectrum at the symmetry point for the undriven Device III appears almost featureless in the measured range of probe frequencies, as can be seen in Fig. 2(f) of the main text. For these reasons we proceed as follows. First, we compare the data of the transmission in the static case with the simulations, using for the dimensionless parameter $\mathcal{N}$ the value $\mathcal{N}=8$ which is somewhat in the center of the estimated range $5\leq\mathcal{N}\leq10$ (see the Methods section of the main text). We do this for different values of the bare frequency $\Delta$, associating to each value of $\Delta$ the coupling $\alpha$ which best reproduces the data. 
Finally, we use the transmission data of the driven device to choose the value of  $\Delta$ that best reproduces, with its associated coupling,  the V-shape of the transmission as a function of pump power and static bias [see Fig.~3(e) of the main text].\\
\indent We note that, independent of the value of $\Delta$ and of the associated coupling $\alpha$, to reproduce the measured levels of transmission in the driven case  we have to double the value of $\mathcal{N}$ in the simulations, with respect to the corresponding static case. Nevertheless, these variations in $\mathcal{N}$ do not affect much the V-shape of the transmission in the pump power-bias plane, which allows to chose the best  value for  $\Delta$.\\
\indent In Fig.~\ref{Sdev3} we compare the measured transmission of the undriven device with simulations performed using different values of $\alpha$. The data used are two perpendicular cuts -- at fixed zero bias and at a fixed probe frequency -- of the experimental colormap in Fig.~2(f) of the main text.  The results are shown for $\Delta/2\pi$ fixed to the value $8$ GHz, namely the one which turns out to give the best agreement with the measurements on the driven device  (the value  used in the main text). The simulations in Fig.~\ref{Sdev3} suggest for Device III the rough estimate  $\alpha=0.8\pm0.1$. 

\clearpage

\begin{figure}[ht!]
\begin{center}
\includegraphics[width=1\textwidth,angle=0]{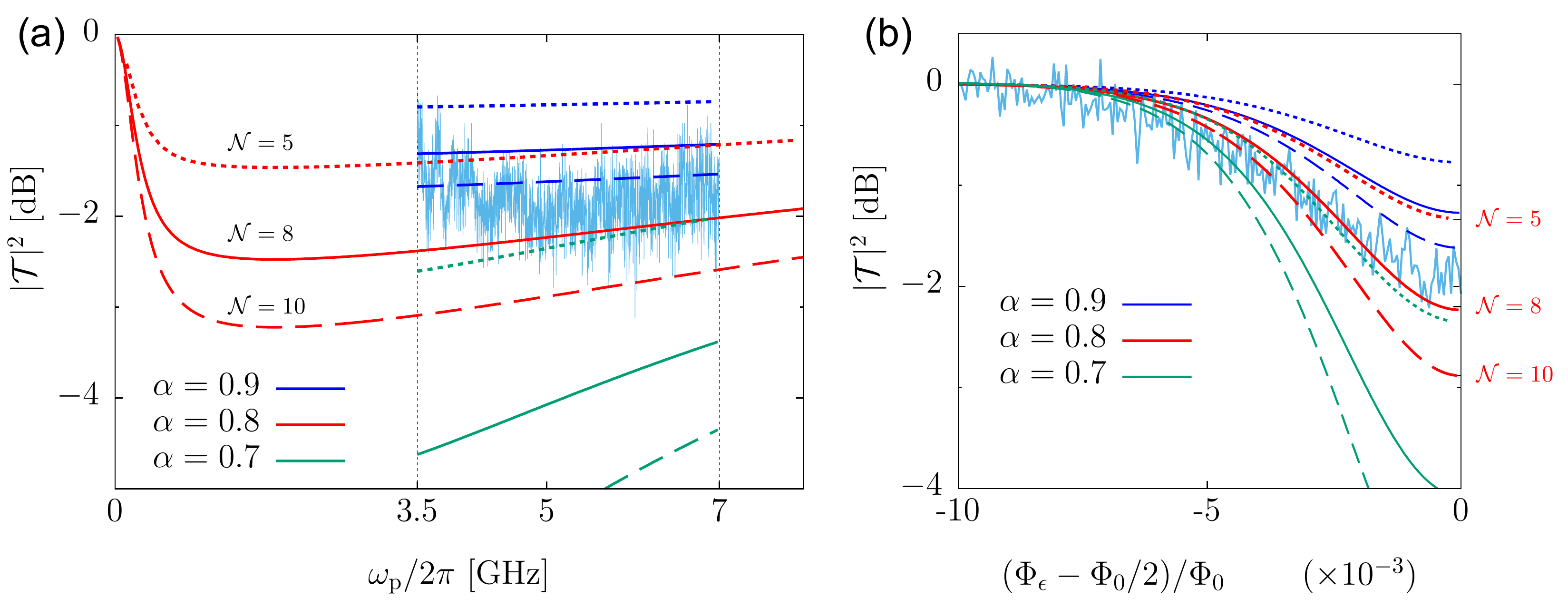}
\caption{\small{Device III -- static case. Measured transmission compared with simulations for three values of the coupling $\alpha$.  (a) --  Transmission as a function of $\omega_{\rm p}$ at the symmetry point (zero bias).  (b) --  Transmission as a function of the static bias with fixed $\omega_{\rm p}=5$~GHz. In both panels, three values of $\mathcal{N}$ are shown for each $\alpha$: $\mathcal{N}=5$ (dotted lines), $\mathcal{N}=8$ (solid lines), $\mathcal{N}=10$ (dashed lines). The plots show that to a larger value of $\mathcal{N}$ there corresponds a larger value of $\alpha$ compatible with the measurements. Simulations are performed by numerically evaluating the   kernels [see Eqs.~(\ref{St-linresp})-(\ref{Skernels})] with the bath correlation function in the exact scaling limit form of Eqs.~(\ref{SQsl1-a})-(\ref{SQsl1-b}). Temperature  and cutoff frequency are  $T=90$ mK and $\omega_{\rm c}/2\pi= 65$ GHz, respectively.}}
\label{Sdev3}
\end{center}
\end{figure}

\subsection{Analytical expression for the qubit's response at weak coupling and zero bias}\label{Suppl8}
At weak coupling, $\alpha\ll 1$, we can approximate the function $\mathcal{W}(x)$, defined in Eq.~(\ref{Sfunctions-c}), as follows 
\begin{eqnarray}\label{SWapprox}
\mathcal{W}(x)=\frac{1}{\alpha+{\rm i}\kappa x}\frac{\Gamma(1+\alpha+{\rm i}\kappa x)}{\Gamma(1-\alpha+{\rm i}\kappa x)}\;
\simeq\;\frac{1}{\alpha+{\rm i}\kappa x}\;.
\end{eqnarray}
Moreover, at zero static bias, $\varepsilon_0=0$, both $H^-$ and $K^-(0)$ vanish. The resulting expression for  the linear susceptibility is 
\begin{eqnarray}\label{Schi-zero-bias}
\chi(\omega_{\rm p})=\frac{H^+(\omega_{\rm p})}{{\rm i}\omega_{\rm p}+K^{+}({\rm i}\omega_{\rm p})}\;.
\end{eqnarray}
Using the linear response expression~(\ref{St-linresp}), which relates $\chi(\omega_{\rm p})$ to the transmission $\mathcal T(\omega_{\rm p})$, we end up with the following approximated expressions
\begin{eqnarray}
\text{Re}\{\mathcal T(\omega_{\rm p})\}&\simeq&1-\mathcal{N}N_-\frac{\kappa\omega_{\rm p} f(\omega_{\rm p})+2N_+(\kappa\omega_{\rm p})^2}{(2\alpha N_+)^2+f^2(\omega_{\rm p})}\;,\label{St-analytical-a}\\
\text{Im}\{\mathcal T(\omega_{\rm p})\}&\simeq&\mathcal{N}\frac{ N_-}{\alpha}\frac{(\kappa\omega_{\rm p})^2 f(\omega_{\rm p})-2\alpha^2 N_+\kappa\omega_{\rm p}}{(2\alpha N_+)^2+f^2(\omega_{\rm p})}\;,\label{St-analytical-b}
\end{eqnarray}
where  $f(\omega_{\rm p})=\alpha^2\omega_{\rm p}+\kappa^2\omega_{\rm p}^3-2 N_+\kappa\omega_{\rm p}$.  \\
\indent  From Eq.~(\ref{Schi-zero-bias}), the imaginary part of the linear susceptibility at zero bias and arbitrary $\alpha$ reads
\begin{eqnarray}\label{SImp1}
\chi''(\omega_{\rm p})=\frac{\text{Im}\{H^+(\omega_{\rm p})\}\text{Re}\{K^+({\rm i}\omega_{\rm p})\}-\text{Re}\{H^+(\omega_{\rm p})\}(\omega_{\rm p}+\text{Im}\{K^+({\rm i}\omega_{\rm p})\})}{\text{Re}\{K^+({\rm i}\omega_{\rm p})\}^2+(\omega_{\rm p}+\text{Im}\{K^+({\rm i}\omega_{\rm p})\})^2}\;.
\end{eqnarray}
Now, for $\alpha\ll 1$, by using the analytical expressions for $H^+$ and $K^+$ with $\mathcal{W}$ in the approximated form given by Eq.~(\ref{SWapprox}), we get 
\begin{eqnarray}
H^+(\omega_{\rm p})&\simeq&\frac{\kappa N_-/\hbar}{\alpha^2+(\kappa\omega_{\rm p})^2} \left(1-{\rm i}\frac{\kappa}{\alpha}\omega_{\rm p}\right)\;,\label{SImp1-b-a}\\
K^+(i\omega_{\rm p})&\simeq&\frac{2\alpha N_+}{\alpha^2+(\kappa\omega_{\rm p})^2} \left(1-{\rm i}\frac{\kappa}{\alpha}\omega_{\rm p}\right)\;,\label{SImp1-b-b}
\end{eqnarray}
 so that the imaginary part of $\chi(\omega_{\rm p})$ acquires the weak coupling form
\begin{eqnarray}\label{SImp1-c}
\chi''(\omega_{\rm p})\simeq-\frac{\kappa N_-}{2\hbar\alpha N_+}\frac{\omega_{\rm p}\text{Re}\{k^+({\rm i}\omega_{\rm p})\}}{\text{Re}\{K^+({\rm i}\omega_{\rm p})\}^2+(\omega_{\rm p}+\text{Im}\{K^+({\rm i}\omega_{\rm p})\})^2}\;.
\end{eqnarray}
In the regime considered here, the peak described by Eq.~(\ref{SImp1-c}) is narrow and the function $\text{Re}\{K^+({\rm i}\omega_{\rm p})\}$ practically constant within its width (roughly measured  by $\text{Re}\{K^+({\rm i}\omega_{\rm p})\}$ itself). The position of the peak is thus well approximated by the value $\omega^*$  obtained upon requiring that $\omega_{\rm p}+\text{Im}\{K^+({\rm i}\omega_{\rm p})\}=0$, which yields
\begin{eqnarray}\label{SOm_star-wc}
\omega^*\simeq \frac{\sqrt{2 N_+ \kappa-\alpha^2}}{\kappa}\;.
\end{eqnarray}
We remark that this approximate analytical result is valid for the unbiased system, $\varepsilon_0=0$, in the limit $\alpha \ll 1$. We did not make use of the above approximated results in the main text. However,  they show how, in the weak coupling regime, the response $\chi''$ acquires a Lorentzian shape. Deviations from this Lorentzian behavior are found as $\alpha$ goes beyond the perturbative regime. This can be seen in Fig.~2 (a-c) of the main text and in Sec.~\ref{Suppl9} below, where a comparison is made of the qubit response in three different coupling regimes which span the range from weak to ultrastrong coupling.

\subsection{Dynamical regimes from transmission for the undriven spin-boson}\label{Suppl9}
\begin{figure}[ht!]
\begin{center}
\includegraphics[width=0.9\textwidth,angle=0]{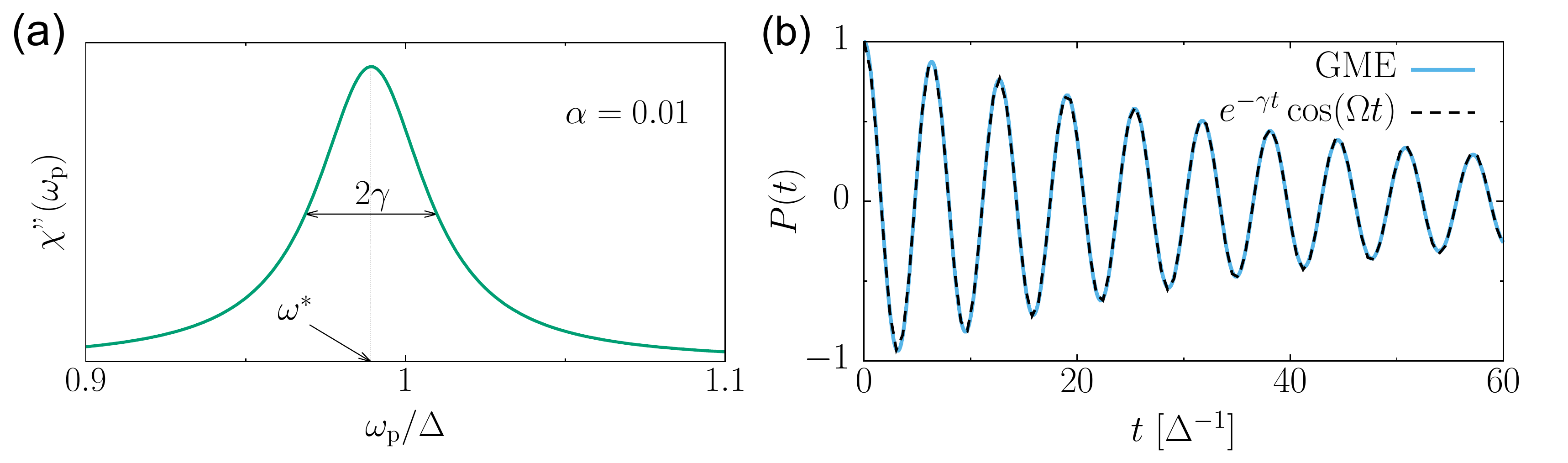}
\caption{\small{Dynamical regime from the susceptibility:  Coherent regime. (a) --  Imaginary part of the linear susceptibility $\chi(\omega_{\rm p})$ (arbitrary units)  calculated by means of the analytical expressions in Eqs.~(\ref{SkapproxSM-a})-(\ref{SkapproxSM-d}).  $\chi''(\omega_{\rm p})$ has a peak at frequency $\omega^*$ with FWHM $2\gamma$. (b) -- Comparison between the dynamics obtained from the GME~(\ref{SGMESM}) (solid line), with $\varepsilon(t)=0$ and bath correlation function $Q(t)$ in the exact scaling limit form of  Eqs.~(\ref{SQsl1-a})-(\ref{SQsl1-b}), and the damped oscillations with renormalized oscillation frequency and decay rate given by $\Omega=\sqrt{(\omega^*)^2-\gamma^2}$ and  $\gamma$, respectively (dashed line). Parameters are $\alpha=0.01$, $T=0.5~\hbar\Delta/k_{\rm B}$, $\varepsilon_0=0$, and $\omega_{\rm c}=10~\Delta$.}}
\label{Sfig:dynamics_weak-coupling}
\end{center}
\end{figure}
\begin{figure}[ht!]
\begin{center}
\includegraphics[width=0.85\textwidth,angle=0]{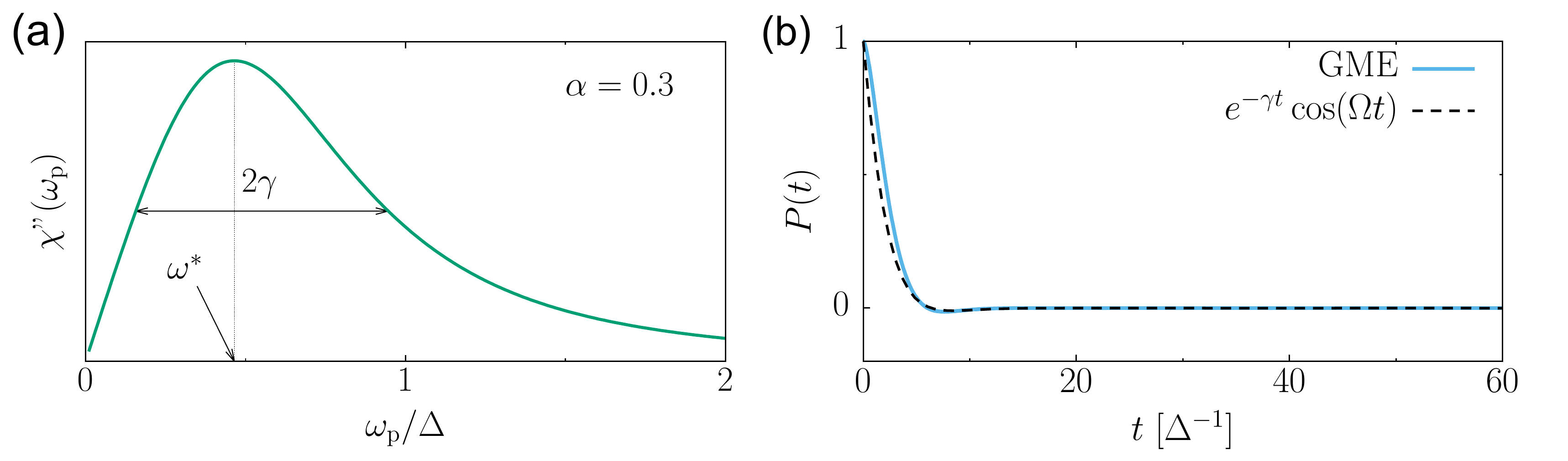}
\caption{\small{Dynamical regime from the susceptibility:  Coherent-incoherent transition regime. (a) --  Imaginary part of linear susceptibility $\chi(\omega_{\rm p})$ (arbitrary units) numerically evaluated by using Eqs.~(\ref{SkpmSM-a})-(\ref{Skernels}) with bath correlation function $Q(t)$ in the exact scaling limit form of  Eqs.~(\ref{SQsl1-a})-(\ref{SQsl1-b}). $\chi''(\omega_{\rm p})$ has  a peak at frequency $\omega^*$  of FWHM $2\gamma$. (b) -- Comparison between the dynamics obtained from the GME~(\ref{SGMESM}) (solid line), with $\varepsilon(t)=0$ and bath correlation function $Q(t)$ in the exact scaling limit form, and the damped oscillations with renormalized oscillation frequency and decay rate given by $\Omega=\sqrt{(\omega^*)^2-\gamma^2}$ and  $\gamma$, respectively (dashed line). Parameters are $\alpha=0.3$, $T=0.5~\hbar\Delta/k_{\rm B}$, $\varepsilon_0=0$, and $\omega_{\rm c}=10~\Delta$.}}
\label{Sdynamics_transition}
\end{center}
\end{figure}
\begin{figure}[ht!]
\begin{center}
\includegraphics[width=0.85\textwidth,angle=0]{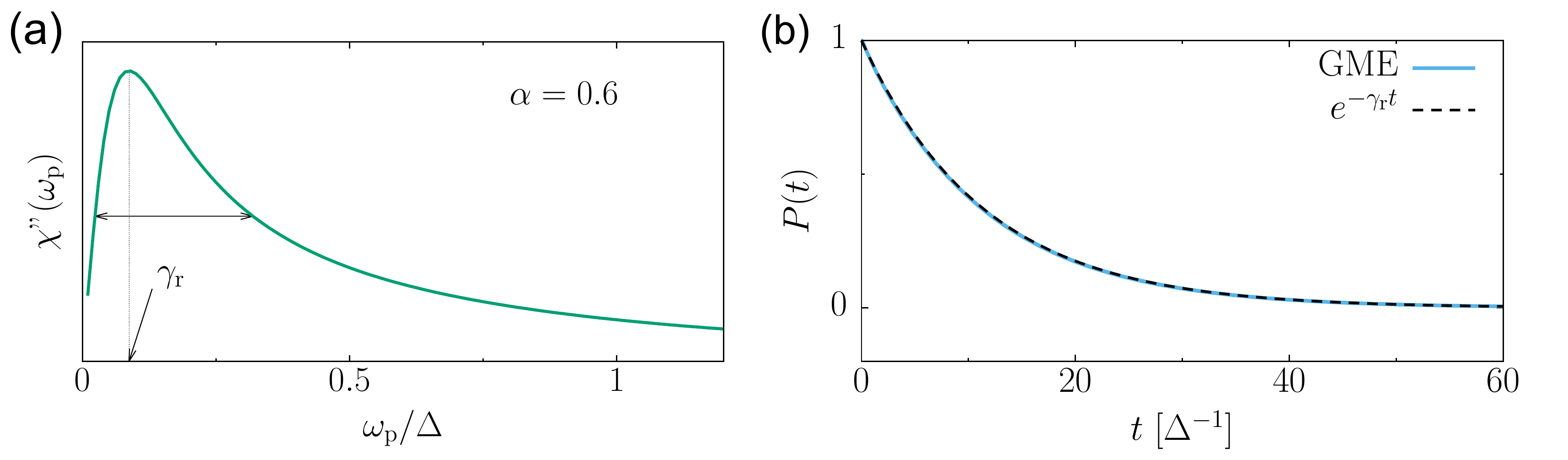}
\caption{\small{Dynamical regime from the susceptibility: Incoherent regime. (a) -- Imaginary part of the linear susceptibility $\chi(\omega_{\rm p})$ (arbitrary units) numerically evaluated by using Eqs.~(\ref{SkpmSM-a})-(\ref{Skernels}) with bath correlation function $Q(t)$ in the exact scaling limit form  of  Eqs.~(\ref{SQsl1-a})-(\ref{SQsl1-b}).  $\chi''(\omega_{\rm p})$ has  a peak at frequency $\gamma_r$. (b) -- Comparison between the dynamics obtained from the GME~(\ref{SGMESM}) (solid line), with $\varepsilon(t)=0$ and bath correlation function $Q(t)$ in the exact scaling limit form, and the exponential decay with rate  $\gamma_r$ (dashed line). Parameters are $\alpha=0.6$, $T=0.5~\hbar\Delta/k_{\rm B}$, $\varepsilon_0=0$, and $\omega_{\rm c}=10~\Delta$.}}
\label{Sfig:dynamics_incoherent}
\end{center}
\end{figure}
In the linear (weak probe) regime, the intrinsic properties of the qubit are not influenced by the presence of the probe field. The dynamical behavior of the qubit in absence of driving is fully encoded in the imaginary part $\chi''$ of the linear susceptibility. Specifically, in the underdamped regime, analogously to the damped harmonic oscillator, the position of the peak of $\chi''$ and its full width at half maximum (FWHM)  are related to the renormalized oscillation frequency and to the decay rate of the oscillations, respectively. Thus, according to Eq.~(\ref{St-linresp}), by measuring the (real part) of the transmission at weak probe, the imaginary part of the susceptibility $\chi$ is accessed which contains the information about the dynamical properties of the static qubit.\\
\indent Consider the case of zero static bias, $\varepsilon_0=0$. The imaginary part $\chi''$ of the susceptibility  is characterized by a peak centered at a frequency $\omega^*$ and of FWHM $2\gamma$. In the coherent regime, occurring when $\omega^*>\gamma$,  the dynamics of $P(t)$ displays damped oscillations with renormalized oscillation frequency $\Omega=\sqrt{(\omega^*)^2-\gamma^2}$ and damping rate $\gamma$. The transition to the incoherent regime is determined by the condition $\omega^*=\gamma$. The incoherent regime, which is realized for $\omega^*<\gamma$, is described by an exponential decay of $P(t)$ with rate $\gamma_r$, the relaxation rate, given in this case by the position of the peak.\\
\indent  As an illustration, let us consider the three different dissipation regimes mentioned above, namely i) coherent, ii) coherent-incoherent transition, and iii) incoherent. We calculate by Eq.~(\ref{SchiSM}) the imaginary part of $\chi$ as a function of the probe frequency and compare the resulting dynamics, namely damped oscillations or incoherent decay with parameters defined by $\omega^*$, $\gamma$, and $\gamma_r$, with the dynamics obtained from direct integration of the GME~(\ref{SGMESM}) in the static, unbiased case, $\varepsilon(t)=0$. Results are shown in Figs.~\ref{Sfig:dynamics_weak-coupling}-\ref{Sfig:dynamics_incoherent}.\\
\indent On the basis of the considerations made above, we are able to establish a phase diagram for the nondriven spin-boson model, i.e., to assign a dynamical behavior (coherent/incoherent) to the points of the coupling-temperature parameter space, by studying $\chi''(\omega_{\rm p})$ and specifically the condition for the coherent-incoherent transition is $\omega^*=\gamma$, where  $\omega^*$ is the position of the peak of  $\chi''(\omega_{\rm p})$ and $2\gamma$ its FWHM.\\
\indent  Such phase diagram, derived within the NIBA, is shown in Fig. 1(c) of the main text for $\omega_{\rm c}=10~\Delta$. The curve, representing the transition temperature $T^*$ as a function of $\alpha$, is an interpolation of the point-set obtained by numerically evaluating $\chi''(\omega_{\rm p})$ by means of Eqs.~(\ref{SkpmSM-a})-(\ref{Skernels}), with the bath correlation function $Q(t)$ in exact scaling limit form in Eqs.~(\ref{SQsl1-a})-(\ref{SQsl1-b}), and searching for the coherent-incoherent transition condition $\omega^*=\gamma$. Specifically, fixing the (dimensionless) temperature to the values $k_{\rm B}T/\hbar\Delta= 2.5,~2,~1.5,~1,~0.75,~0.5,~0.25,$ and $0.1$, a numerical search for the value of $\alpha$ realizing the condition  $\omega^*=\gamma$ was performed.
The lowest point, of abscissa $\alpha=0.5$, is individuated by the exact result $k_{\rm B} T^*(\alpha=0.5)/\hbar\Delta=\Delta/2\omega_{\rm c}$~\cite{Weiss2012}.

\end{document}